\documentclass[preprint2]{aastex}
\usepackage{color}
\usepackage{epsfig}
\defcitealias{jeffery93a}{JH93}

\slugcomment{Draft version May 2017.}

\shorttitle{Chemical compositions of V652\,Her and HD\,144941}
\shortauthors{Pandey \& Lambert}

\begin{document}

\title{Non-Local Thermodynamic Equilibrium abundance analyses of the extreme helium stars: V652\,Her and HD\,144941.}

\author{Gajendra Pandey}
\affil{Indian Institute of Astrophysics;
Bangalore, 560034 India}
\email{pandey@iiap.res.in}
\and
\author{David L.\ Lambert}
\affil{The W.J. McDonald Observatory and Department of Astronomy, University of Texas at Austin; Austin,
TX 78712-1083}
\email{dll@astro.as.utexas.edu}

\begin{abstract}

 Optical high-resolution spectra of V652\,Her and HD\,144941, the two extreme helium stars with 
exceptionally low C/He ratios, have been subjected to a non-LTE abundance analysis using the tools 
TLUSTY and SYNSPEC. Defining atmospheric parameters were obtained from a grid of non-LTE atmospheres 
and a variety of spectroscopic indicators including He\,{\sc i} and He\,{\sc ii} line profiles, 
ionization equilibrium of ion pairs such as C\,{\sc ii}/C\,{\sc iii} and N\,{\sc ii}/N\,{\sc iii}. 
The various indicators provide a consistent set of atmospheric parameters: 
$T_{\rm eff}$=25000$\pm$300K, $\log g$ = 3.10$\pm$0.12(cgs), and $\xi=13\pm2 {\rm km\,s^{-1}}$ are
provided for V652\,Her, and $T_{\rm eff}$=22000$\pm$600K, $\log g$ = 3.45$\pm$0.15 (cgs), 
and $\xi=10 {\rm km\,s^{-1}}$ are provided for HD\,144941. In contrast to the non-LTE analyses, 
the LTE analyses -- LTE atmospheres and a LTE line analysis -- with the available indicators do not 
provide a consistent set of atmospheric parameters. The principal non-LTE effect on the elemental 
abundances is on the neon abundance. It is generally considered that these extreme helium stars with 
their very low C/He ratio result from the merger of two helium white dwarfs. Indeed, the derived composition 
of V652\,Her is in excellent agreement with predictions by Zhang \& Jeffery (2012) who model the slow merger 
of helium white dwarfs; a slow merger results in the merged star having the composition of the accreted white dwarf.  
In the case of HD\,144941 which appears to have evolved from metal-poor stars a slow merger is incompatible with 
the observed composition but variations of the merger rate may account for the observed composition.  
More detailed theoretical studies of the merger of a pair of helium white dwarfs are to be encouraged.

\end{abstract}

\clearpage

\section{Introduction}

Extreme helium stars (EHes) are very hydrogen poor stars with effective temperatures 
of about 10000 K to 30000 K (i.e., spectral types A and B) with surface gravities of 
$\log g \sim 1$ for the coolest stars and increasing to $\sim 3$ for the hottest stars. 
A majority of EHes populate a locus of roughly constant $log L/M \sim 4.5$ where the 
luminosity $L$ and mass $M$ are in solar units. This locus most likely represents an 
evolutionary track with stars evolving at about constant luminosity from low to high temperatures. Such EHes thought 
to form from the merger of a helium white dwarf with a carbon-oxygen white dwarf have 
carbon-to-helium ratios by number of about 0.6 \% with presently analyzed stars exhibiting 
C/He ratios by number in the range 0.3 \% to 1.0 \%. The carbon is provided by the surface 
of the C-O white dwarf and the helium primarily by the helium white dwarf.

At the time of  \citet{jeffery08.hdef3.b}'s succinct review of EHes, 
about 20 Galactic EHes were known. Two are set apart from the majority highlighted above 
by a much lower 
C/He ratio. V652\,Her according to \citet{jeffery01b} and an LTE analysis 
has the low C/He ratio of 0.006 \% and HD\,144941 according to \citet{harrison97}'s 
LTE abundance analysis has the even lower C/He ratio of 0.0017 \%. Both ratios are 
sharply lower than the C/He ratio of the majority of the EHes. In addition, V652\,Her 
and HD\,144941 have higher surface gravities than the majority EHes of the same 
effective temperature and thus correspond to a $\log L/M$ smaller by about a factor of 1.3 dex than the majority.
These differences especially the low C/He ratio suggest a different origin, namely the merger 
of a helium white dwarf with another helium white dwarf. Again, see \citet{jeffery08.hdef3.b}'s 
review for further details.

In this paper, we describe a non-LTE analysis of new high-quality optical spectra of both 
V652\,Her and HD\,144941 primarily in order to determine how non-LTE effects influence the 
C/He ratio but also to measure the effects of departures from LTE on the abundances of other 
elements. The paper follows our similar analyses of non-LTE effects on several EHes
having C/He ratios characteristic of the majority  EHes \citep{pandey11,pandey14}.

\section{Observations}

 High-resolution optical spectra of V652\,Her and HD\,144941
were obtained on 2011 May 13 at the coud\'e focus of the 
W.J. McDonald Observatory's Harlan J. Smith 2.7-m telescope
with the Robert G. Tull cross-dispersed \'echelle spectrograph
\citep{tull95} at a resolving power of R = 60,000. Three
thirty minutes exposures were recorded for each of these stars.
The observing procedure and the wavelength coverage is same as 
described in \citet{pandey01}.
The Image Reduction and Analysis Facility (IRAF) software package 
was used to reduce these recorded spectra. HD\,144941's spectrum obtained
from Anglo-Australian Telescope (AAT), and analysed by \citet{przybilla05},
was made available by Simon Jeffery (private communication) for comparison. 

The sample wavelength interval shown in Figure 1  displays
the extracted spectrum from each exposure of the observed EHes.
All spectra were aligned to the rest wavelengths of well-known lines. 
Inspection of the Figure 1 shows that the line profiles are not always symmetric.
Note the obvious asymmetry in the two exposures of V652\,Her attributable to
atmospheric pulsations \citep{jeffery01b}. V652\,Her pulsates with a period of
2.592 hours \citep{landolt75} and  a radial velocity amplitude of about 
70 km s$^{-1}$ \citep{hill81}. For the abundance analysis, we have used the spectrum of V652\,Her,
showing symmetric profiles with signal-to-noise ratio of about 140 per pixel at 5600\AA. 
The line profiles of HD\,144941 for each exposure appear symmetric and show extremely weak metal lines. 
Hence, these exposures were coadded to enhance the signal-to-noise ratio for 
the abundance analyses; the signal-to-noise ratio is about 280 per pixel at 5600\AA.

The pure absorption line spectrum of V652\,Her 
is dominated by contributions from the following species: H\,{\sc i}, He\,{\sc i}, N\,{\sc ii},
N\,{\sc iii}, O\,{\sc ii}, Ne\,{\sc i}, Al\,{\sc iii}, Si\,{\sc ii},
Si\,{\sc iii}, S\,{\sc ii}, S\,{\sc iii}, and Fe\,{\sc iii}.
However, the absorption line spectrum of HD\,144941 is dominated by 
H\,{\sc i}, and He\,{\sc i} lines and a small collection of weak lines from other species.
Revised Multiplet Table (RMT) \citep{moore72}, tables of spectra
of H, C, N, and O \citep{moore93} and the {\it NIST} Atomic Spectra
Database\footnote{http://www.nist.gov/pml/data/asd.cfm} (ver. 5.3) were used for line identification. 

The primary objective is to determine reliable atmospheric parameters (effective temperature, surface gravity 
and microturbulence) and then the chemical composition of V652\,Her and HD\,144941.

\section{Quantitative Fine Analyses}

Non-LTE line-blanketed model atmospheres are used to determine the atmospheric parameters
and the chemical composition. The effective temperature $T_{\rm eff}$ and 
surface gravity $g$ are obtained from  intersecting loci in the $T_{\rm eff}$ 
versus $\log g$ plane. These loci represent the ionization equilibrium of 
available ion pairs such as 
C\,{\sc ii}/C\,{\sc iii}, N\,{\sc ii}/N\,{\sc iii}, Si\,{\sc ii}/Si\,{\sc iii}, Si\,{\sc iii}/Si\,{\sc iv} and
S\,{\sc ii}/S\,{\sc iii} and loci  derived from the best fits to the
Stark-broadened profiles of He\,{\sc i} and He\,{\sc ii} lines. 
The microturbulent velocity $\xi$ is obtained from both  the N\,{\sc ii} and
the O\,{\sc ii} lines with each ion providing lines spanning a range in equivalent width.

In principle, the chemical composition of the adopted model atmosphere
must match the composition derived from a spectrum. This match is achieved
iteratively. Note that model atmospheres computed with
C/He of 0.003 - 0.03\% and H/He of 0.0001 and 0.1 have the same
atmospheric structure and so provide the same atmospheric parameters including the
composition. This helpful aide to the abundance analysis arises because neutral He is
the dominant opacity source for the two EHes.

Since  photoionization of 
neutral helium is the main source of continuous opacity,
lines of another species, say C\,{\sc ii}, are sensitive to the
C/He abundance ratio. Abundances are given as $\log\epsilon$(X)
and normalized with respect to $\log\Sigma\mu_{\rm X}\epsilon$(X) $=$ 12.15
where $\mu_{\rm X}$ is the atomic weight of element X. Since all elements but He have a
very low abundance, the logarithmic He abundance is 11.54.

Our abundance analyses were carried with non-LTE model atmospheres and non-LTE
(and LTE) line formation for all major elements. For a few  minor elements, the abundance
analysis could be done only in LTE. 
Partially line-blanketed non-LTE model atmospheres were computed with the 
code TLUSTY \citep{hubeny88,hubeny95} using atomic data and model atoms
provided on the TLUSTY home page\footnote{http://nova.astro.umd.edu/index.html}, 
as described in \citet{pandey14}. These model atmospheres included opacity from
both bound-free and 
bound-bound transitions of H, He, C, N, O, Ne, Mg, Si, S, and Fe in
non-LTE. The adopted model atoms with their number of levels given in parentheses, are:
H\,{\sc i}(9), He\,{\sc i}(14), He\,{\sc ii}(14), C\,{\sc i}(8), C\,{\sc ii}(11), C\,{\sc iii}(12),
C\,{\sc iv}(13), N\,{\sc i}(13), N\,{\sc ii}(6), N\,{\sc iii}(11), N\,{\sc iv}(12), O\,{\sc i}(22),
O\,{\sc ii}(29), O\,{\sc iii}(29), Ne\,{\sc i}(35), Ne\,{\sc ii}(32), Ne\,{\sc iii}(34),
Mg\,{\sc ii}(14), Si\,{\sc ii}(16), Si\,{\sc iii}(12), Si\,{\sc iv}(13), S\,{\sc ii}(14),
S\,{\sc iii}(20), Fe\,{\sc ii}(36), and Fe\,{\sc iii}(50). Of the model atoms available in TLUSTY,
these choices each refer to the smallest of the available atoms for many ions. Use of these model 
atoms suffices for the calculations of model atmospheres but, as we describe below, larger model atoms 
are used for the calculations of equivalent widths of lines.

Model atmospheres in LTE were also computed using TLUSTY.  Model grid in non-LTE and LTE
were computed covering the ranges $T_{\rm eff} = 20\,000 (1\,000) 30\,000$ K 
and $\log g = 3.0 (0.1) 4.5$ cgs.

In this paper, we have used TLUSTY and SYNSPEC for calculating LTE and
non-LTE model atmospheres and line profiles \citep{hubeny88,hubeny94,hubeny95,hubeny2011,hubeny2017}.
Stellar atmospheric parameters provided in Table 1 are
determined from the spectra of V652\,Her and HD\,144941 on the assumption
of both non-LTE and LTE using the lines given in Tables 2 and 3.
Except where noted, the $gf$-values of the lines are taken from the
NIST database\footnote{http://www.nist.gov/pml/data/asd.cfm} (ver. 5.3).
A few other sources consulted for $gf$-values are given in footnotes to the
relevant tables.

TLUSTY model atmospheres are calculated with line opacity provided by the smallest of
the available model atoms for many ions, as detailed above. In many cases, the observed lines are
not contained within these model atoms. In order to extend the non-NLTE calculations to more of
the observed lines, we ran the statistical equilibrium calculations with larger model atoms
available in TLUSTY as described in \citet{hubeny2017}. Results are provided for the following model atoms:
C\,{\sc ii}(22), C\,{\sc iii}(23), N\,{\sc ii}(42), N\,{\sc iii}(32), O\,{\sc ii}(29),
Ne\,{\sc i}(35), and  Si\,{\sc iii}(30). Lines with both the upper and lower level within the 
model atom are marked by * in Table 2 and 3.
We note that, our non-LTE abundance analysis using model atmospheres with extended (more levels) model atoms is
fairly consistent in terms of the ionization balance and the derived abundances when compared with
the analysis using small (less levels) model atoms; the mean abundances differ by
about 0.05 to 0.3 dex, and that individual line abundances differ by typically $\leq$0.3 dex. 
However, for most of the ions, we notice that the larger model atoms
provide relatively less line-to-line scatter in the derived abundances.

For the elements H to Fe, identification of lines suitable for analysis
is not a major issue for V652\,Her. In particular, a good selection of clean lines
representing the ions of key elements is available. For HD\,144941, lines of H, He, N, and O are available in good number but
 fewer lines are present for C, Ne, Al, Si, and S with Fe represented only by upper limits to equivalent widths of the most promising lines of Fe\,{\sc iii}.
\citet{moore93} is the primary source of wavelengths and
classifications for these lines.

\subsection{V652\,Her}

\subsubsection{Non-LTE analyses}

The Non-LTE code SYNSPEC \citep{hubeny94} was adopted to compute the line profiles and the
theoretical equivalent widths using the non-LTE model atmospheres. The observed absorption profile or 
its measured equivalent width was matched with the SYNSPEC prediction to obtain the 
non-LTE abundance. The unresolved blends of two or more lines were dealt by
synthesizing and then matching to the observed feature by adjustment of abundances.

For determining the $T_{\rm eff}$, $\log g$ and $\xi$, a standard procedure is followed.
The microturbulent velocity $\xi$ is estimated from N\,{\sc ii} and O\,{\sc ii} lines as they 
show a wide range in equivalent width. To minimise the temperature dependence, N\,{\sc ii} lines 
with similar lower excitation potentials (LEP) were used: N\,{\sc ii} lines were used with LEPs 
about 18, 21, 23, and 25 eV. $\xi$ was found from the requirement that the derived abundance is
independent of the measured equivalent width. A microturbulent velocity 
$\xi=13\pm2 {\rm km\,s^{-1}}$ is obtained from N\,{\sc ii} and O\,{\sc ii} lines.

Ionization equilibrium is imposed, using model atmospheres computed with small model atoms, to provide loci in 
the ($T_{\rm eff}$, $\log g$) plane for the following pairs of ions:
C\,{\sc ii}/C\,{\sc iii}, N\,{\sc ii}/N\,{\sc iii}, S\,{\sc ii}/S\,{\sc iii}, 
and Si\,{\sc iii}/Si\,{\sc iv} but low weight is given to the last ratio because the Si abundances from the Si\,{\sc iii} lines show
a large line-to-line scatter. 

Fits to the Stark-broadened wings of He\,{\sc i} and He\,{\sc ii} line profiles provide additional loci
in the ($T_{\rm eff}$, $\log g$) plane. The line broadening coefficients are from TLUSTY/SYNSPEC which
adopts the line broadening coefficients for He\,{\sc i} 4471\AA, 4388\AA, and 4026\AA\ 
from \citet{barnard1974,shamey1969}. For He\,{\sc i} 4009\AA, TLUSTY/SYNSPEC uses an approximate Stark broadening
treatment. For He\,{\sc ii} 4686\AA, TLUSTY/SYNSPEC uses a broadening table from 
Schoening (private communication to I. Hubeny). The predicted line profiles depend on the
electron densities and, therefore, on the temperature and surface gravity.
Observed profiles of the He\,{\sc i} 4471\AA, 4388\AA, 4026\AA\
and 4009\AA\ lines and the He\,{\sc ii} 4686\AA\ were used in the analysis.

Sample observed profiles of the He\,{\sc i} and He\,{\sc ii} 4686\AA\
line are shown in Figure 2 with predicted non-LTE profiles for a non-LTE
atmosphere of $T_{\rm eff}$=25000K and three different surface gravities. The He\,{\sc i} and He\,{\sc ii}
loci were obtained by fitting the line profiles for a range of effective temperatures.
Note that, the predicted profiles have been convolved with the instrumental profile
and the stellar rotation profile. A projected rotation velocity of 10-12 km s$^{-1}$ was obtained
from fits of synthetic spectra to clean O\,{\sc ii} lines with an allowance for the instrumental
profile. 

The loci derived from the application of ionization equilibrium to C, N, Si, and S ions are
also added to the loci from the He\,{\sc i} and He\,{\sc ii} profiles.
Figure 3 shows these loci. Their intersection suggests the best non-LTE model
atmosphere has $T_{\rm eff}$=25000$\pm$300K and $\log g$ = 3.10$\pm$0.12.

H\,{\sc i} observed profiles at 4102\AA, 4340\AA\ and 4861\AA\ were chosen for
estimating the non-LTE hydrogen abundance by spectrum synthesis.
The line wings of 4340\AA\ and 4861\AA\
profiles are mainly used for this purpose. It is noted that the best fitting theoretical profile for each line
does not have an emission core but
such cores appear for theoretical profiles of higher H abundance.
Less weight is given to the NLTE hydrogen abundance derived from the poor signal-to-noise profile
of 4102\AA\ line. The hydrogen model atoms and the
line broadening coefficients adopted from \citet{VCS} are from TLUSTY.
Observed profiles of the 4102\AA, 4340\AA, and 4861\AA\
are shown in Figure 4 with predicted non-LTE profiles for a non-LTE
atmosphere of $T_{\rm eff}$=25000K and $\log g$ = 3.10 for three different
hydrogen abundances.

The abundances of all elements were derived for the adopted model atmosphere
($T_{\rm eff}$, $\log g$, $\xi$)=(25000, 3.10, 13.0), computed with extended model atoms. The final photospheric line by line
non-LTE abundances including the mean abundance
and the line-to-line scatter are given in Table 2.
The abundance rms errors due to uncertainty in $T_{\rm eff}$ and $\log g$, from
C\,{\sc ii}, C\,{\sc iii}, N\,{\sc ii}, N\,{\sc iii}, O\,{\sc ii},
Ne\,{\sc i}, Mg\,{\sc ii}, Si\,{\sc iii}, Si\,{\sc iv}, S\,{\sc ii}, S\,{\sc iii}, and
Fe\,{\sc iii} are 0.03, 0.13, 0.03, 0.16, 0.04, 0.02, 0.04, 0.05, 0.13, 0.05, 0.05, and
0.03 dex, respectively.

\subsubsection{LTE analyses}

Analysis of the line spectrum was repeated with the TLUSTY LTE models and LTE line
analysis. This LTE analysis uncovers several inconsistencies arising from substantial
non-LTE effects on some of the lines. These inconsistencies demand worrying compromises
in selecting the atmospheric parameters and, thus, in determining the elemental abundances.

Among the concerns are the fits to the He\,{\sc i} and He\,{\sc ii} line profiles. Observed profiles of He\,{\sc i} and the 
He\,{\sc ii} 4686 \AA\ lines are shown in Figure 5 with predicted LTE profiles for the LTE 
atmosphere of $T_{\rm eff}$ = 25300 K and three different surface gravities. The predicted 
He\,{\sc i} profiles fail to reproduce the observed cores but are acceptable fits to 
the line wings. Note that the non-LTE profiles provide a satisfactory fit  both to  the cores and 
the wings (Figure 2). In LTE, the fit to the He\,{\sc ii} 4686 \AA\ line requires a much higher 
surface gravity at a given temperature than other indicators. Figure 6, the LTE counterpart to Figure 3,
shows the He\,{\sc i} and He\,{\sc ii} loci, as well as those corresponding to ionization equilibrium. 

The loci set by ionization equilibrium for C, N and Si are almost coincident and each is only slightly shifted from
their non-LTE location in the $T_{\rm eff},\log g$ plane. However, the locus set by S\,{\sc ii}/S\,{\sc iii} is shifted away
from other ionization equilibrium loci because of the large non-LTE effect on the S\,{\sc iii} lines.

Other species subject to appreciable non-LTE effects do not enter into consideration in determining the
atmospheric parameters. The species most obviously affected by non-LTE effects is Ne\,{\sc i} .

An enthusiast dedicated to LTE analyses with Nelsonian eyesight might adopt the LTE model (see Figure 6)
with $T_{\rm eff} = 25300\pm300$ K,  $\log g = 3.25\pm0.12$ and a microturbulence of 13 km s$^{-1}$.
The LTE abundances for this adopted LTE TLUSTY model are given in Table 2. The abundance rms errors,
due to uncertainty in $T_{\rm eff}$ and $\log g$, from Al\,{\sc iii}, Si\,{\sc ii}, P\,{\sc iii},
and Ar\,{\sc ii} are 0.05, 0.07, 0.03, and 0.04 dex, respectively. The abundance rms errors for the
rest of the species are very similar to those estimated for the appropriate non-LTE model atmosphere. Of course,
such errors do not recognize that the choice of the LTE model atmosphere involves compromises.

\subsection{HD\,144941}

\subsubsection{Non-LTE analyses}

Essentials of the  procedure  discussed in Section 3 for V652\,Her were adopted for the non-LTE analyses of HD\,144941.
A microturbulent velocity $\xi=10 {\rm km\,s^{-1}}$ is used as suggested by \citet{harrison97}.
Except for the observed H\,{\sc i} and He\,{\sc i} lines, all lines in the observed spectrum are
weak.  Of course, the derived abundances from these weak lines are almost independent of the adopted
microturbulence.

Fits to the Stark-broadened wings of He\,{\sc i} and He\,{\sc ii} line profiles provide loci
in the ($T_{\rm eff}$, $\log g$) plane. Unfortunately, the He\,{\sc ii} 4686\AA\ line profile is not
detected on our spectrum but the upper limit to its presence provides a limiting locus 
in the ($T_{\rm eff}$, $\log g$) plane. Figure 7 shows the
sample observed profiles with predicted non-LTE profiles for a non-LTE atmosphere of 
$T_{\rm eff}$=22000K and three different surface gravities.

The only locus obtained from ionization balance is through sulphur ions:
S\,{\sc ii}/S\,{\sc iii} but this is based on just one weak S\,{\sc ii} and two weak S\,{\sc iii} lines. 

Figure 8 shows the loci obtained from the fits to the He\,{\sc i} and He\,{\sc ii} profiles
and the ionization balance of S\,{\sc ii}/S\,{\sc iii} using model atmospheres computed with small model atoms. 
These intersecting loci are used in determining 
the final model parameters of $T_{\rm eff}$=22000$\pm$600K and 
$\log g$ = 3.45$\pm$0.15.

Observed profiles of the 4102\AA, 4340\AA, and 4861\AA\ are shown in Figure 9 with predicted
non-LTE profiles for a non-LTE atmosphere of $T_{\rm eff}$=22000K and $\log g$ = 3.45 for three different
hydrogen abundances. The line-wings are mainly used for this purpose as the predicted profiles show 
emission in the line-core. Unlike V652\,Her, the best fits to the wings are 
affected by emission in the core with the intensity of emission increasing from H$\delta$ to H$\beta$.
However, the predicted profiles by \citet{przybilla05} match the 
observations (the wings as well as the core) fairly well.
 For comparison, observed profiles of the 4102\AA, 4340\AA, and 4861\AA\ lines for three different H abundances
are shown in Figure 10 with predicted
non-LTE profiles for a non-LTE atmosphere with the stellar parameters, $T_{\rm eff}$=22000K and $\log g$ = 4.15
adopted by \citet{przybilla05}: 
($\log \epsilon(\rm H)$ = 10.1 is the non-LTE abundance estimated by \citet{przybilla05} and \citet{przybilla06}.


The non-LTE abundances for the adopted non-LTE TLUSTY model 
($T_{\rm eff}$, $\log g$, $\xi$)=(22000, 3.45, 10.0), computed with small model atoms, are given in Table 3.
The abundance rms errors, due to uncertainty in $T_{\rm eff}$ and $\log g$, from
C\,{\sc ii}, N\,{\sc ii}, O\,{\sc ii}, Ne\,{\sc i}, Mg\,{\sc ii}, Si\,{\sc iii}, S\,{\sc ii}, S\,{\sc iii},
and Fe\,{\sc iii} are 0.06, 0.06, 0.05, 0.04, 0.07, 0.06, 0.08, 0.06, and 0.04 dex, respectively.

\subsubsection{LTE analyses}

Inconsistencies among atmospheric parameter indicators when using LTE model atmospheres and
LTE line analysis techniques may be expected to resemble those inconsistencies
identified as present for V652\,Her. The lower abundances of many elements in 
HD\,144941 relative to V652\,Her may effect the radiation
field in HD\,144941's atmosphere even though opacity at many wavelengths including the optical
region is dominated by helium.

Observed profiles of the He\,{\sc i} and He\,{\sc ii} 4686\AA\
line are shown in Figure 11 with predicted LTE profiles for a LTE
atmosphere of $T_{\rm eff}$=21000K and three different surface gravities.
As anticipated, the predicted LTE He\,{\sc i} profiles for HD\,144941
fail to reproduce the observed line-core but the line-wings are well reproduced but the predicted
non-LTE He\,{\sc i} profiles successfully reproduce the observed core as well as the wings 
with the adopted non-LTE TLUSTY model (see Figure 7).

The He\,{\sc ii} 4686\AA\ line is not
positively detected on our spectra. Predicted profiles of this line provide a limiting locus with significantly
higher gravities than other indicators, a similar situation occurs for V652\,Her (Figure 6).

Figure 12 shows the loci obtained from the fits to the He\,{\sc i} and He\,{\sc ii} profiles
and ionization equilibria which are provided by the the following pairs of ions:
Si\,{\sc ii}/Si\,{\sc iii} and S\,{\sc ii}/S\,{\sc iii}. The Si locus appears in Figure 12 but not Figure 8 because
TLUSTY lacks an adequate model Si$^+$ atom.
The final compromise LTE model parameters are 
$T_{\rm eff}$=21000$\pm$600K and $\log g$ = 3.35$\pm$0.15.

LTE abundances for the adopted LTE model are given in Table 3.
Observed profiles of the Balmer lines 4102\AA, 4340\AA, and 4861\AA\ are shown in Figure 13 with predicted
LTE profiles for a LTE atmosphere of $T_{\rm eff}$=21000K and $\log g$ = 3.35 for three different
hydrogen abundances. The abundance rms errors,
due to uncertainty in $T_{\rm eff}$ and $\log g$, from Al\,{\sc iii}, and Si\,{\sc ii}
are 0.06, and 0.11 dex, respectively. The abundance rms errors for the
rest of the species are very similar to those estimated for the appropriate non-LTE model atmosphere.
In comparing non-LTE $-$ LTE abundance differences in Table 2 for V652\,Her and in Table 3 for
HD\,144941, it must be noted that the compromise LTE model for V652\,Her is 300 K hotter and
0.15 dex greater in $\log g$ than its non-LTE model but the LTE model for HD\,144941 is 1000 K cooler
and 0.10 dex lower in $\log g$ than its non-LTE model.

\section{Discussion $-$ Chemical Composition}

Tables 4 and 5 summarize our derived non-LTE and LTE abundances for V652\,Her and HD\,144941, respectively.
Mean elemental abundances are given for elements represented by more than a single stage of
ionisation. Composition of the solar photosphere is given 
in the final column \citep{asplund09}.

\subsection{V652\,Her}

Inspection of the abundances for V652\,Her offers three pointers to the star's
history: i) most obviously, it is  H-poor by a factor of about 300, ii) CNO-cycling was
most likely responsible for conversion of H to He because, as first noted by \citet{jeffery99},
the star is N-rich and relatively C and O poor, and iii) the overall metallicity of the
star is approximately solar, as judged by the abundances of elements from Mg to Fe. Before
connecting these points to the star's evolutionary status, brief remarks are made on the previous LTE abundance analysis
of this star.

\citet{jeffery01b} obtained a time series of optical spectra at a resolving power of 10000.
Results of a LTE abundance analysis when this pulsating star was near maximum radius were
given. This LTE atmosphere had parameters $T_{\rm eff} = 22000$ K, $\log g = 3.25$ and
a microturbulence of 9 km s$^{-1}$. This model is 3000 K cooler than our LTE model and differs slightly in
surface gravity and microturbulence. A direct comparison of LTE abundances (our Table 4 and their Table 2)
gives differences of less than $\pm0.3$ dex for all elements except C, Ne and P for which differences in the
sense (us $-$ them) are $-0.4$, $+0.6$ and $-0.9$ dex, respectively. Jeffery et al. isolate their P abundance
estimate for comment and recommend that `The older value [as provided by \citet{jeffery99}]
should be preferred for the present.' This older value for P is within 0.1 dex of our LTE value. 
When the 1999 values for H to Fe are adopted, the (us $-$ them) differences are within $\pm0.3$ dex except
for S at 0.4 dex. The 1999 LTE model atmospheres corresponded to
$T_{\rm eff} = 24550\pm500$ K, $\log g = 3.68\pm0.05$ and a microturbulence 5 km s$^{-1}$. In short, 
our and published LTE abundance analyses are in good agreement but this should be hardly surprising given
the similarities of spectra and analytical tools.


 \citet{przybilla06} provided a hybrid non-LTE analysis (i.e., a non-LTE analysis of absorption lines was
made using a LTE model atmosphere) of a selection of lines measured off the spectrum used by \citet{jeffery01b}.
The atmosphere was very similar to that used by \citet{jeffery01b} and, thus, 3000 K cooler  but of similar surface
gravity to our chosen non-LTE model atmosphere. The abundances of H, C, N, O, Mg and S differ in the
sense (Us $-$ Prz) by $+0.3, -0.4, +0.2, -0.5, -0.8$ and $-0.3$ dex, respectively. C, Mg and S were represented 
by very few lines. With the exception of the Mg abundance from the 
Mg\,{\sc ii} 4481 \AA\ feature (the sole Mg indicator), the non-LTE $-$ LTE
corrections are within  $\pm0.2$ dex. This independent non-LTE analysis fully confirms the result 
that V652\,Her's atmosphere is now highly enriched in CNO-cycled material.

What is new here is the demonstration that adoption of a set of non-LTE atmospheres and the
chosen collection of non-LTE line tools reveal inconsistencies in the 
indicators previously employed to determine the
appropriate atmospheric parameters and corrections to LTE abundances for non-LTE effects which
can for some species (e.g., Ne\,{\sc i}) be considerable. At the present time, the non-LTE abundances
should be used to discuss the three pointers mentioned in this section's opening paragraph.
   
Even if V652\,Her is the result of a merger of two He white dwarfs, it is very likely to have
retained the metallicity - say, Ne to Fe abundances - of the two stars from which the white
dwarfs evolved with mass loss and mass exchange playing major roles. For the five elements
(Ne, Mg, Si, S and Fe) with non-LTE abundances, the mean difference (V652\,Her $-$ Sun) is $-0.1$ which
suggests a near-solar initial composition. (The differences of $-0.5$ for Mg and Fe are intriguing.) 
Thus, V652\,Her is a member of the thin disk.

V652\,Her's high N abundance is 0.8 dex above the solar value. This coupled with the high H deficiency
and sub-solar values of C and O point to CNO-cycling as the primary process for conversion of
H to He. The CNO-cycles preserve the total number of C, N and O nuclei. The non-LTE
CNO-sum is 8.78 and the solar sum is 8.92. Almost fortuitously, the expected sum for a mix corresponding
to a metal deficiency of $-0.1$ dex is 8.82!.  


\subsection{HD\,144941}

Comparison of the non-LTE and solar abundances (Table 5) shows a near-uniform difference (star $-$ Sun)
for elements from C to Fe. With the exception of Ne and S, the mean difference is $-1.6$ with individual
values ranging from $-1.4$ to $-1.8$.  (The LTE Al abundance gives a difference of $-1.4$.)
The difference for Fe is $\leq -0.9$. The Fe line at 5156\AA\ is present in all  three exposures, and clearly seen in Figure 14 but since this is
the sole positive detection and not confirmed by other lines,  we assign an upper limit to the Fe abundance. 
Adopting the Fe abundance
from {\it IUE} spectra and the LTE analysis  by \citet{jeffery97}, the difference
is $-1.8$, a value consistent with our much higher limit. Their logarithmic Fe abundance of $5.7\pm0.2$ and our
non-LTE abundances, except for the $\alpha$-elements: Ne and S, are consistent with the composition of metal-poor stars, 
residents of the Galactic thick disk or halo, which show $\alpha$-element enhancements of approximately
$+0.3$ for O, Mg and Si. Moreover, the abundances of C, N and O are consistent with the difference
of $-1.6$, and, hence, the H-deficiency is not obviously attributable to CNO-cycling, as is the case for
V652\,Her.

As just noted, Ne and S provide striking exceptions to the run of (star $-$ Sun) differences offered by
other elements. These elements give differences of $-0.7$ for Ne and $-0.9$ for S. There are no
convincing nucleosynthetic reasons for these differences. (Just conceivably, N-rich He-rich
material could have been heated and a considerable amount for N burnt to Ne by two
successive $alpha$-captures. The total initial sum of CNO abundances could have been about
7.3 but the Ne abundance of 7.2 implies a near perfect conversion of CNO to Ne.) The Ne\,{\sc i}
lines appear to be secure identification - see Figure 14 for the strongest Ne\,{\sc i} line. The primary
suspect for the high S abundance are systematic errors associated with the identification of
the weak S\,{\sc ii} and S\,{\sc iii} lines. The gf-values and non-LTE treatment are less likely
sources of errors as the measured lines are among a larger set used for V652\,Her. 

Our LTE abundances are in good agreement with the previous LTE analysis of an optical spectrum
by \citet{harrison97} using the model atmosphere $T_{\rm eff} = 23200$ K, $\log g = 3.9$ and
a microturbulence of 10 km s$^{-1}$.  The abundance differences (us $-$ them) are within  $\pm0.25$
dex limits. Unfortunately, Harrison \& Jeffery did not include Ne and S. Their Fe abundance of 6.4 was
lowered to 5.7 by their synthesis of lines in {\it IUE} spectra \citep{jeffery97}. Our non-LTE abundances are also in
good agreement but for Mg with results from \citet{przybilla06}'s hybrid non-LTE analysis: differences (Us $-$ Pry) are 
$+0.3$, $0.0$, $-0.3$, $-0.4$ and $-0.8$ dex for H, C, N, O and Mg, respectively. Our corrections (Non-LTE $-$ LTE) are in 
agreement to within $\pm0.15$ dex including for Mg\,{\sc ii} with those found by \citet{przybilla06}. The model atmosphere
used by \citet{przybilla06} has the same effective temperature but a higher surface gravity ($\log g = 4.15$ rather our 3.45). 
\citet{przybilla06} selected lines from the optical spectrum used by \citet{harrison97}. 

\section{Concluding Remarks}

A possible origin for extreme He stars such as V652\,Her and HD\,144941 involves, as noted in the Introduction, the merger of two He
white dwarfs. The white dwarfs themselves began life as main sequence stars in a binary system which then experienced two mass  exchange
and mass loss episodes  to form the pair of low-mass He white dwarfs. Emission of gravitational radiation causes the white dwarfs to slowly approach
each other.  White dwarf binaries which can merge in a Hubble time or less are candidates to account for V652\,Her and HD\,144941.
\citet{zhang12a} present evolutionary tracks for the merger of two white dwarfs. Three modes of merger are discussed: slow, fast and
composite.  Since their calculations are most extensive for $Z=0.02$ white dwarfs,  V652\,Her is discussed first.

In the slow merger (see also \citet{saio00}), the lower mass white dwarf loses its mass in a few minutes to  form a disk around the more
massive white dwarf.  Accretion from the disk by the surviving white dwarf may last several million years and completes the merger. The composite
star is initially a luminous red giant  which evolves to become a hot subdwarf before entering the white dwarf cooling track.  Along this evolutionary
track, the star may appear as an EHe.  Zhang \& Jeffery's calculations predict that the surface composition of the merged star (i.e., the EHe) is that of the less massive white dwarf; there is no mixing between the accreted material and the accreting star. The recipe for setting the compositions of the He white dwarfs as described by Zhang \& Jeffery proves insensitive to the less massive star's assumed mass (see Figure 2 from \citet{zhang12a} for $Z = 0.02$ star). The stars are predicted to be N-rich (N/C $\sim 100$ and N/O $\sim 10$) with C/He $\sim 0.01\%$. This pattern for He, C, N, O and also Ne matches to within a factor of about two, the observed composition of V652\,Her (Table 4). (Oddly, the predicted mass fraction of $^{24}$Mg is about an order
of magnitude higher than observed.) Predicted evolutionary tracks in the ($T_{\rm  eff}, \log g)$ plane about the EHe domain are coincident for
masses of 0.5-0.7$M_\odot$ with tracks for higher masses occurring at lower surface gravities (see Figure 15 from 
\citet{zhang12a}). (The tracks
at $T_{\rm eff} \leq 40000 K$ appear insensitive to the initial composition; $Z= 0.02$ was generally adopted, i.e., slightly supra-solar.) With
non-LTE parameters from Table 1, V652\,Her falls on the predicted $0.8M_\odot$ track which is only 0.6 dex lower in $\log g$ than the 0.5-0.7$M_\odot$
tracks.  Since \citet{jeffery01b} estimate V652\,Her's mass at $0.6\pm0.2M_\odot$, we consider the observed star fits the predicted evolutionary track
for a slow merger of two He white dwarfs. This fit is echoed by the correspondence between predicted and observed compositions.  In short, V652\,Her
may be the result of a slow merger of two low-mass He white dwarfs.

In a fast merger, accretion is considered to be complete in a few minutes. In Zhang \& Jeffery's simulations the envelope resembles a hot
corona with carbon produced by He-burning and N destroyed by $^{14}$N($\alpha,\gamma)^{18}$O and at higher temperatures $\alpha$-capture
converts the $^{18}$O to $^{22}$Ne. Burning in a He-shell occurs in flashes. Zhang \& Jeffery conclude description of fast mergers with
the remark that `For all the fast merger models the surface composition is rich in  $^{12}$C, $^{18}$O and $^{22}$Ne but there is almost
no $^{14}$N for models with initial composition of $Z = 0.02$. (Of course, it is frustrating that isotopic wavelength shifts for atomic lines with 
stellar line widths do not permit determinations of the isotopic mix of C, O and Ne.) For slow mergers with $Z = 0.02$, the predicted N/C ratio is about 100 and independent of the final mass as noted above, but for fast mergers the N/C is predicted to run from 0.01 for a final mass of $0.5M_\odot$ to $10^{-9}$ for a final mass of $0.8M_\odot$ (see Figure 19 of Zhang \& Jeffery).  Clearly V652\,Her is not the result of a fast merger.

In the composite model, as envisaged by Zhang \& Jeffery, the first phase of accretion transfers about half of the mass as a fast merger
with the second half transferred via a disk as in the case of a slow merger. For mergers resulting in a total mass of less than about 0.6$M_\odot$
(apparently, resulting from roughly equal masses for the two white dwarfs), the surface convection zone is absent and, thus, the surface
composition resulting from accretion from the disk is that of the accreted white dwarf, i.e., the predicted composition is equivalent to that of a slow
merger (see Figure 20 from Zhang \& Jeffery). At total masses above $0.6M_\odot$ the predicted compositions tend in the direction to those predicted by a
fast merger; the N/C ratio is about 0.3 for a mass 0.8$M_\odot$ but the prediction for a fast merger is $10^{-11}$. V652\,Her may
have resulted from a composite merger (effectively, a slow merger) with a total mass less than about $0.6M_\odot$. 
 
In brief, V652\,Her's  composition and location in the $(T_{\rm eff},\log g)$ plane encourage the idea that the star resulted from the
merger of two Helium white dwarfs. As modeled by Zhang \& Jeffery,  the star resulted from a slow or a composite merger with a rather
relaxed constraint on the masses of the white dwarfs. A fast merger, as defined by Zhang \& Jeffery, leads to a star with N/C less than
unity in sharp conflict with the observed N/C ratio of 50. Significantly, V652\,Her's composition and position in the $(T_{\rm eff},\log g)$ plane
can be met by a range of slow or composite mergers. In contrast to V652\,Her, the picture of a merger of He white dwarfs is not so
obviously readily applicable to HD\,144941. 

An initial impression of HD\,144941's composition (Table 5) is that the C, N and O are consistent with the initial abundances for a star
with [Fe/H] of about $-1.6$ or $Z \sim 0.0004$ but such a consistency surely sits uneasily with the messy conversion of a main sequence star to a highly-evolved H-poor star. Moreover, the star's Ne abundance is about 1 dex greater than anticipated for a normal metal-poor star. Thus, we have sought possible solutions in Zhang \& Jeffery's paper. It is worthy of note that with respect to the $(T_{\rm eff},\log g)$ plane, HD\,144941's non-LTE parameters provide an excellent fit to Zhang \& Jeffery's evolutionary tracks for merging white dwarfs from stars with $Z=0.001$. Observed and predicted compositions are more difficult to reconcile. For slow mergers at $Z = 0.001$, models predict products with a much lower C/He ratio than observed and a very high N/C ratio (see Figure 20 from Zhang \& Jeffery), say N/C $\sim 200$ but the observed ratio is N/C $\sim 0.3$. Fast mergers invert the N/C ratio and but are likely to provide a N abundance declining steeply with increasing total mass such that a match to HD\,144941 will be found only for a narrow range of masses. A composite merger may match the observed He, C and N abundances with adjustments to Zhang \& Jeffery's recipe for this merger process. Introduction of a period of
fast merging is expected to increase the Ne (as $^{22}$Ne) abundance and so possibly match the observed Ne abundance. (An episode of fast merging
may also provide abundant $^{18}$O.) Expansion of the parameter space considered by Zhang \& Jeffery is highly desirable.

Our abundance analysis shows that the use of non-LTE effects in the construction of model atmospheres and analyses of absorption lines for He-rich warm stars leads to significant changes in the defining atmospheric abundances and in certain elemental abundances relative to the assumption of LTE for
model atmospheres and abundance analysis. Judged by our determinations of composition and location in the $(T_{\rm eff},\log g)$ plane, V652\,Her  is likely to have resulted from the merger of two helium white dwarfs. HD\,144941's location in the  $(T_{\rm eff},\log g)$ plane is similarly consistent with the merger hypothesis. Its composition may also be consistent with formation through a merger but additional theoretical predictions seem required. These conclusions are based on Zhang \& Jeffery's bold and exploratory calculations of the merging process.  Perhaps, our analyses will not only encourage refinements to the study of white dwarf mergers but also a search for additional EHe stars with the very low C/He ratio that is a characteristic feature of V652\,Her and HD\,144941.

Quite fortuitously, after submission of the paper,  \citet{jefferyMN2017} reported the discovery of a third EHe star with a low C/He (= 0.0023\%) ratio. The star
{\it GALEX} J184559.8-413827 according to the LTE abundance analysis by \citet{jefferyMN2017}  has a subsolar ($-0.4$)  metallicity with  C, N and O  abundances
similar to that of V652\,Her, i.e., the atmosphere of the new discovery is rich in CNO-cycled material. Our abundance analyses of V652\,Her  show that a non-LTE
reanalysis of J184559.8-413827 will not alter this conclusion.

\acknowledgments
We thank Simon Jeffery and Mike Montgomery for helpful email exchanges. We also thank
Ivan Hubeny for helping us in using the TLUSTY and SYNSPEC codes. We would like to thank 
the anonymous referee for the constructive comments.
DLL acknowledges the support of the Robert A. Welch Foundation of Houston, Texas through grant F-634.

\begin{figure}
\epsscale{1.00}
\plotone{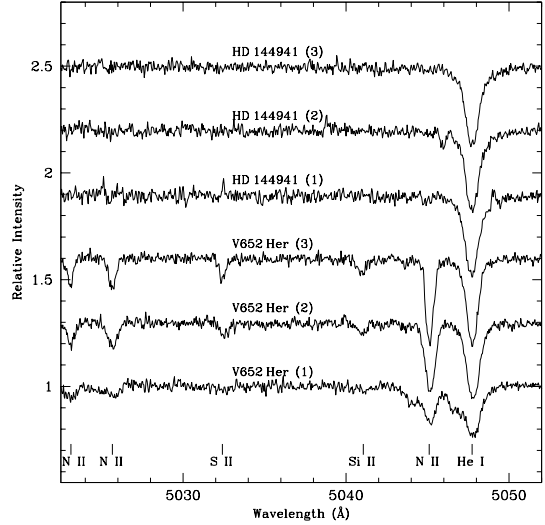}
\caption{Sample spectral region is shown for three exposures each of V652\,Her and HD\,144941.
The exposure numbers are given in parentheses, and the positions of the key lines are 
identified in this window from 5023 to 5052 \AA.}
\end{figure}

\begin{figure}
\epsscale{1.00}
\plotone{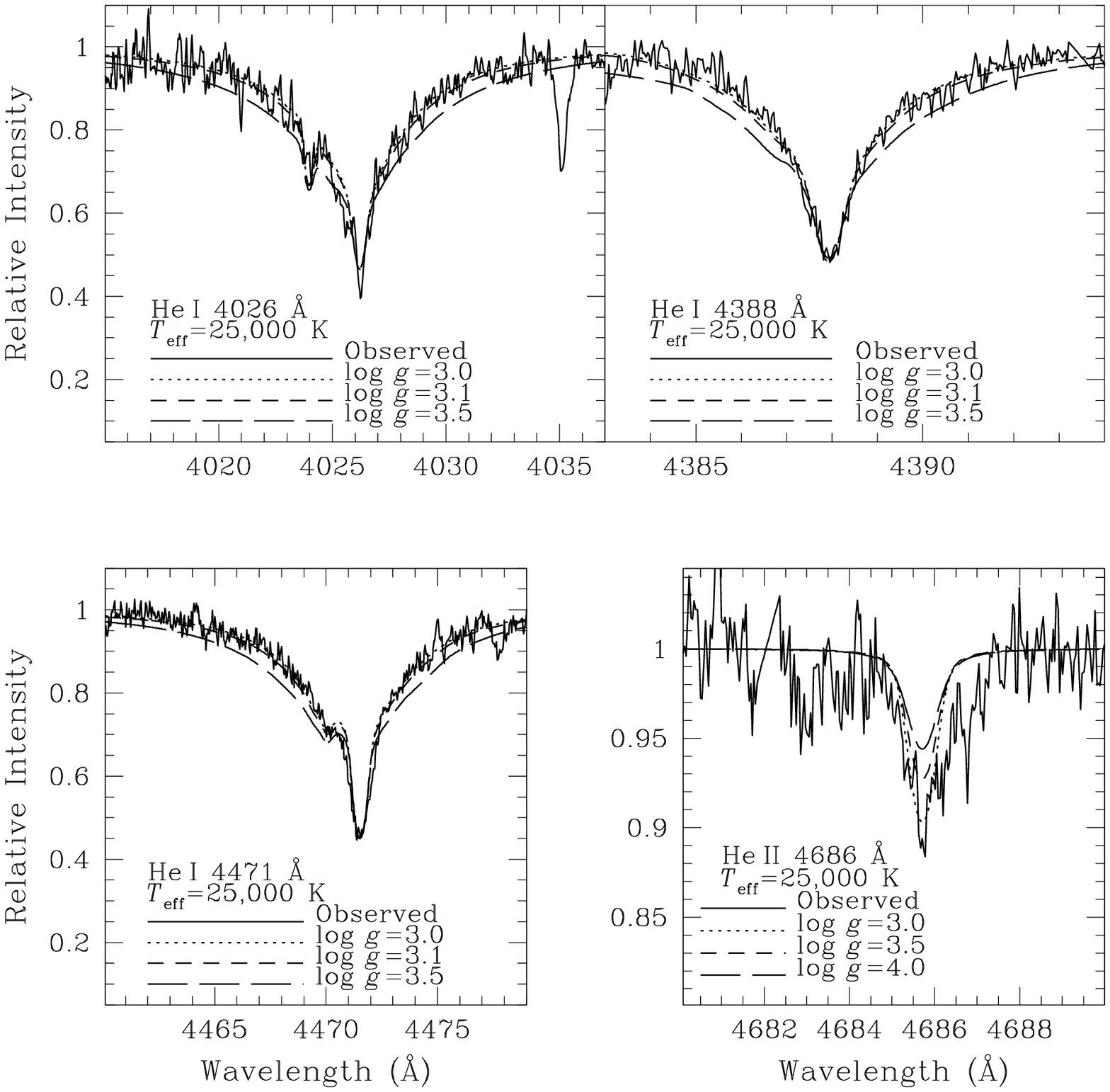}
\caption{The observed spectrum of V652\,Her and theoretical NLTE He\,{\sc i} and
He\,{\sc ii} line profiles calculated using the
NLTE model $T_{\rm eff}$=25,000 K, for three different $\log g$ values $-$ see key
on the figure.}
\end{figure}

\begin{figure}
\epsscale{1.00}
\plotone{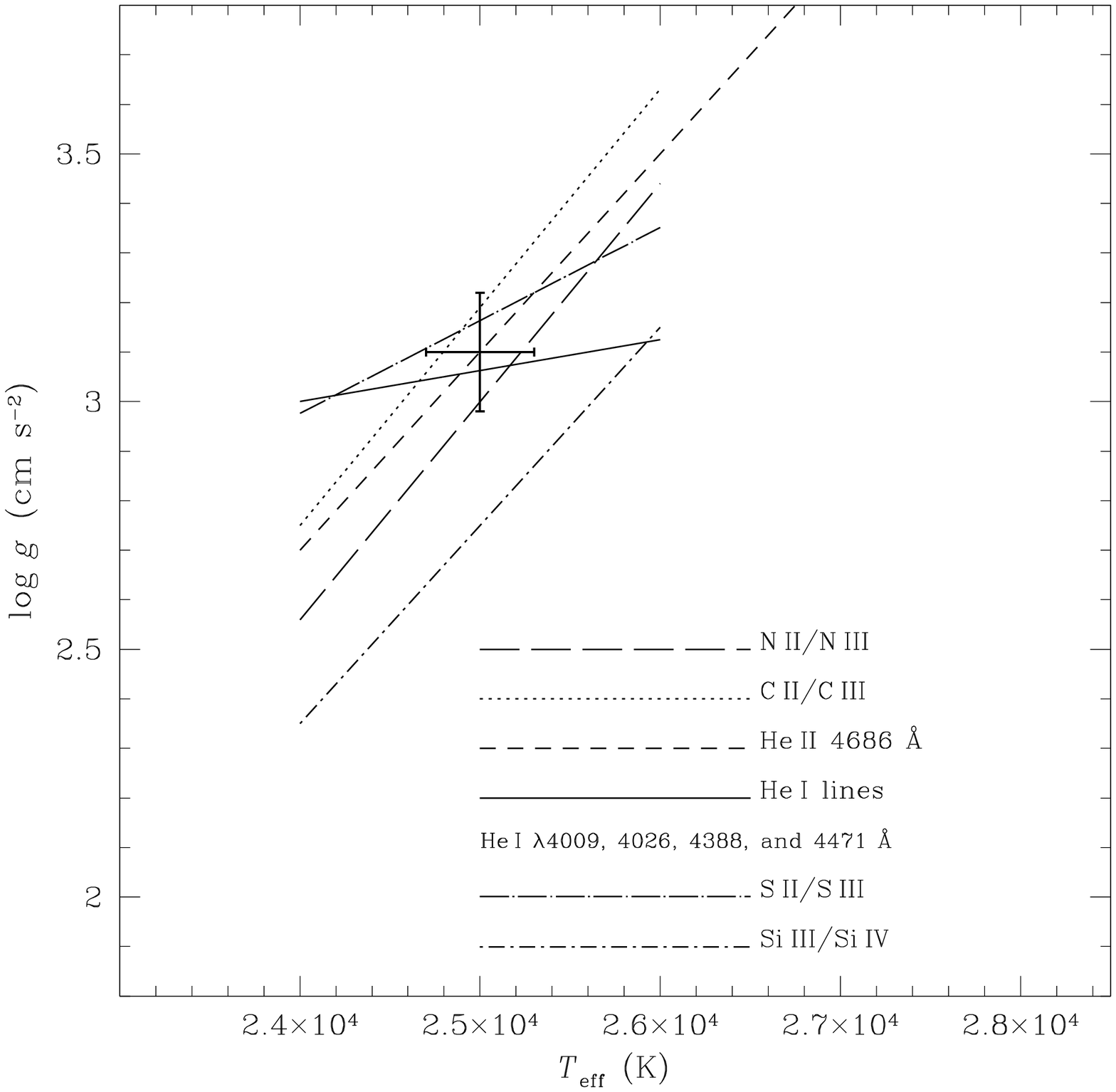}
\caption{The $T_{\rm eff}$ vs $\log g$ plane for 
V652\,Her.
Loci satisfying ionization equilibria are plotted $-$ see keys on the figure.
The loci satisfying optical He\,{\sc i} and He\,{\sc ii} line profiles
are shown. The cross shows the adopted 
NLTE model atmosphere parameters.}
\end{figure}

\begin{figure}
\epsscale{1.00}
\plotone{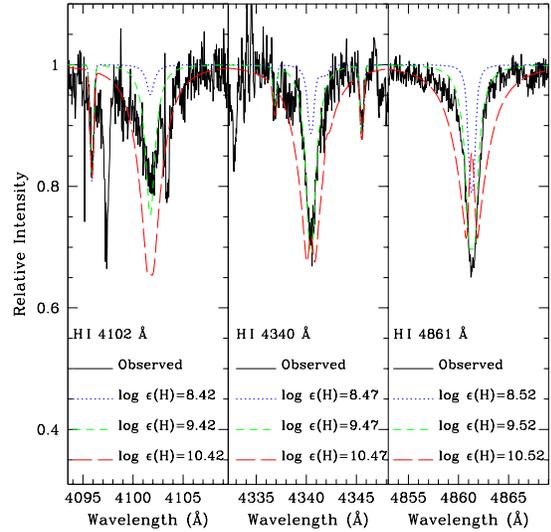}
\caption{The observed spectrum of V652\,Her and theoretical NLTE H\,{\sc i} line
profiles calculated using the
NLTE model $T_{\rm eff}$=25,000 K and $\log g$ = 3.1, for three different
H abundances $-$ see key on the figure.}
\end{figure}

\begin{figure}
\epsscale{1.00}
\plotone{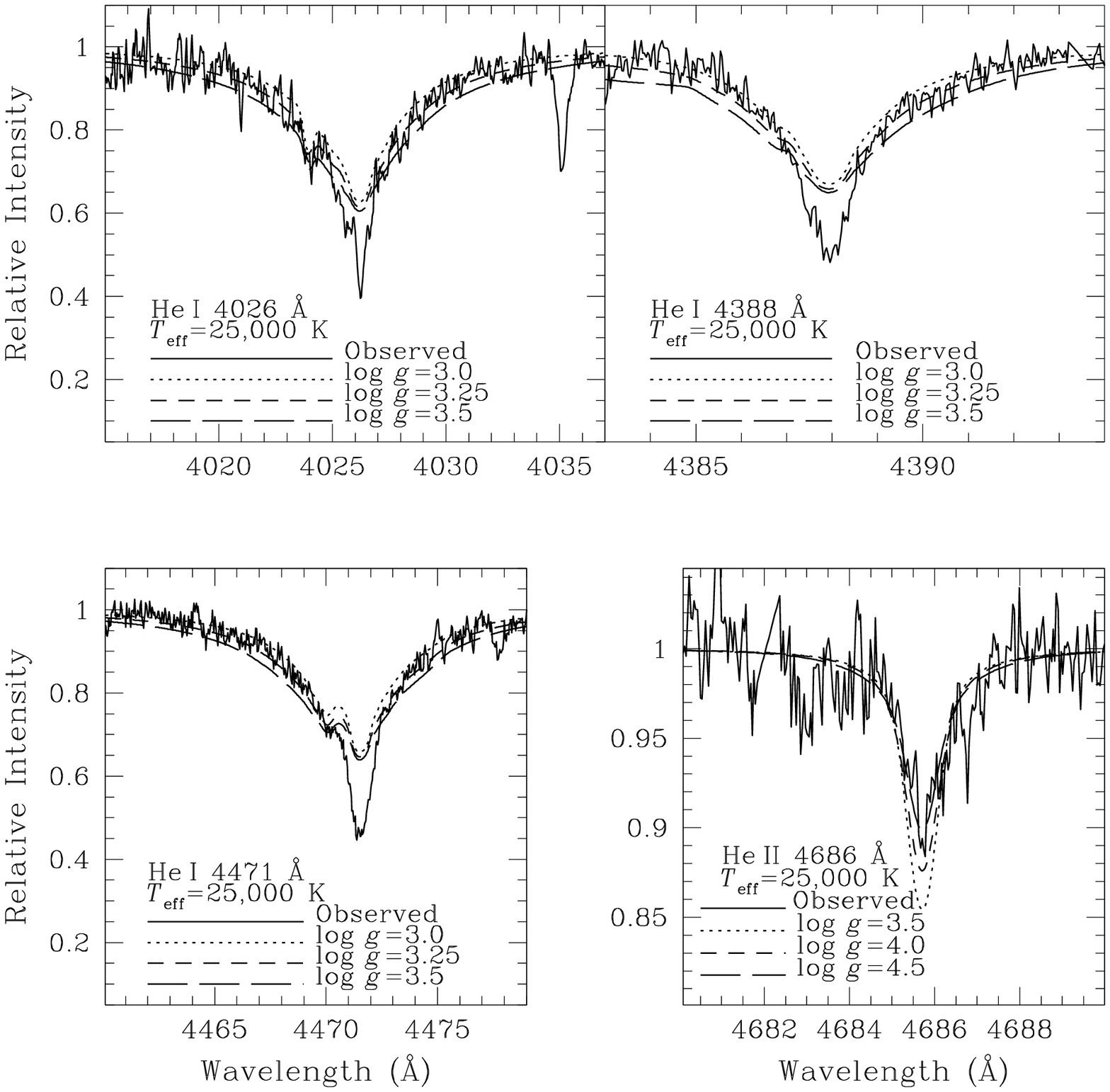}
\caption{The observed spectrum of V652\,Her and theoretical LTE He\,{\sc i} and
He\,{\sc ii} line profiles calculated using the
LTE model $T_{\rm eff}$=25,300 K, for three different $\log g$ values $-$ see key
on the figure.}
\end{figure}

\begin{figure}
\epsscale{1.00}
\plotone{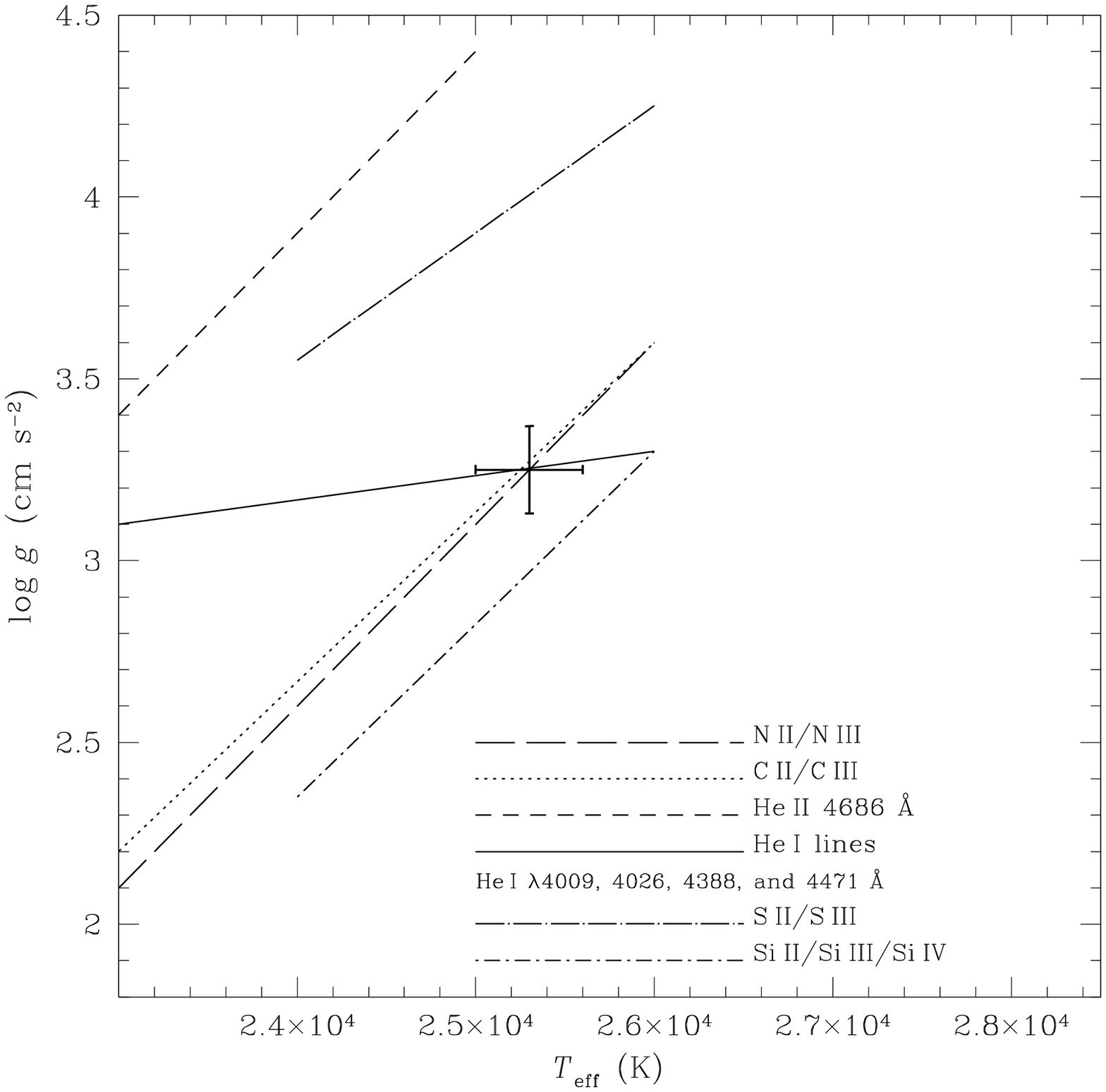}
\caption{The $T_{\rm eff}$ vs $\log g$ plane for
V652\,Her.
Loci satisfying ionization equilibria are plotted $-$ see keys on the figure.
The loci satisfying optical He\,{\sc i} and He\,{\sc ii} line profiles
are shown. The cross shows the adopted
LTE model atmosphere parameters.}
\end{figure}

\begin{figure}
\epsscale{1.00}
\plotone{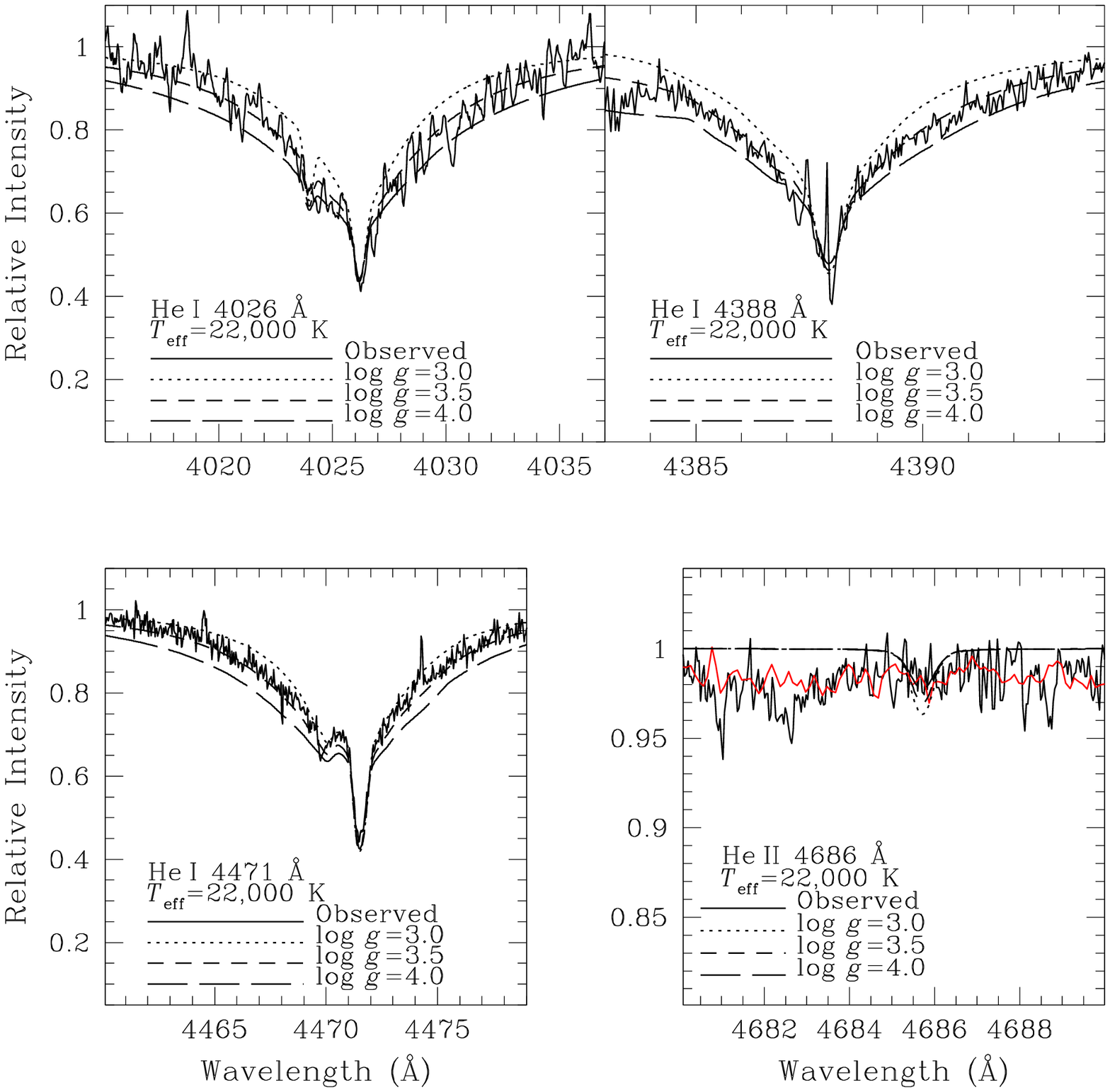}
\caption{The observed spectrum of HD\,144941 and theoretical NLTE He\,{\sc i} and
He\,{\sc ii} line profiles calculated using the
NLTE model $T_{\rm eff}$=22,000 K, for three different $\log g$ values $-$ see key
on the figure. The AAT spectrum is shown in red.}
\end{figure}

\begin{figure}
\epsscale{1.00}
\plotone{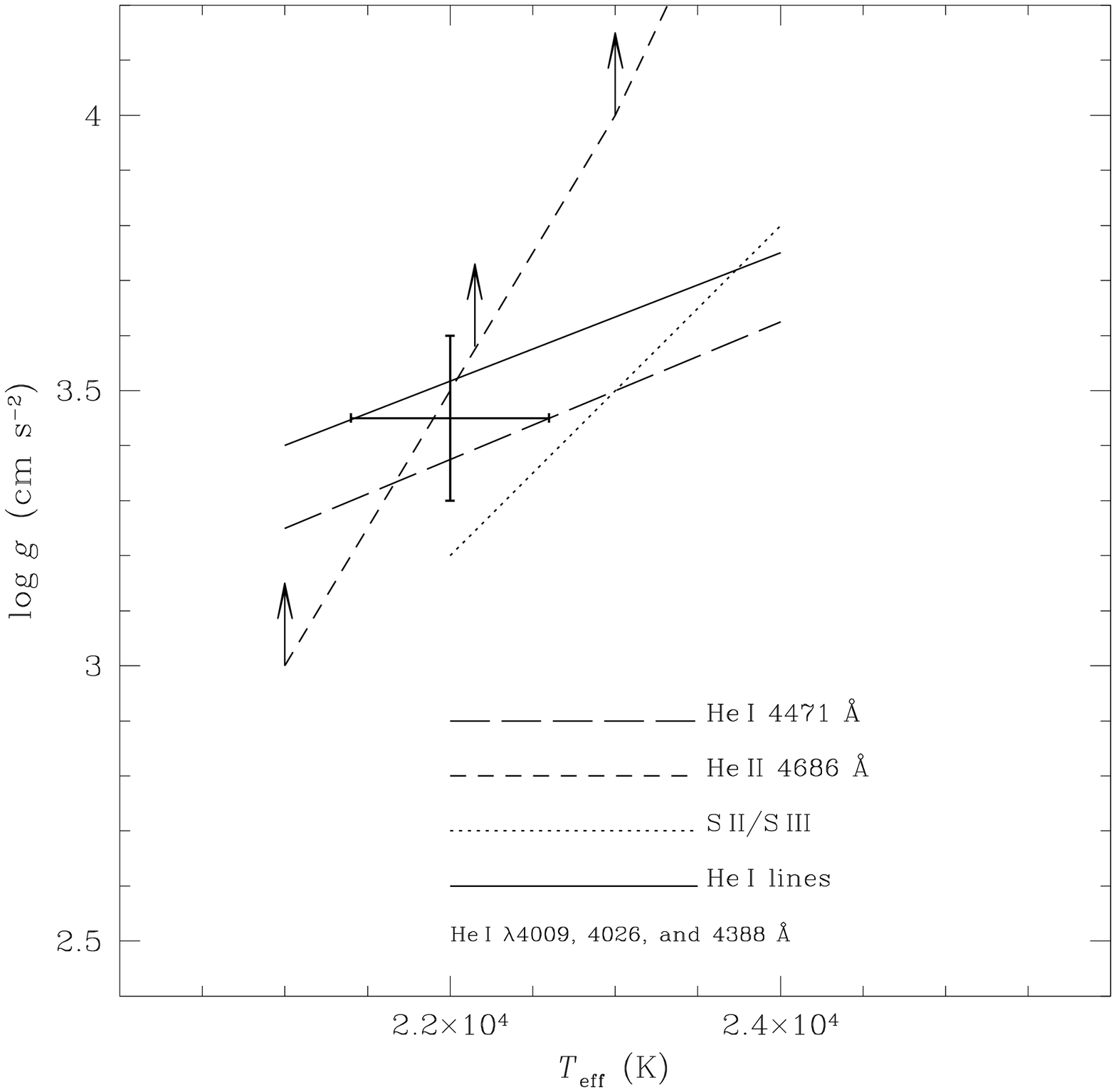}
\caption{The $T_{\rm eff}$ vs $\log g$ plane for HD\,144941.
Loci satisfying ionization equilibria are plotted $-$ see keys on the figure.
The loci satisfying optical He\,{\sc i} line profiles are shown. 
The limiting locus for He\,{\sc ii} is shown by upward arrows.
The cross shows the adopted NLTE model atmosphere parameters.}
\end{figure}

\begin{figure}
\epsscale{1.00}
\plotone{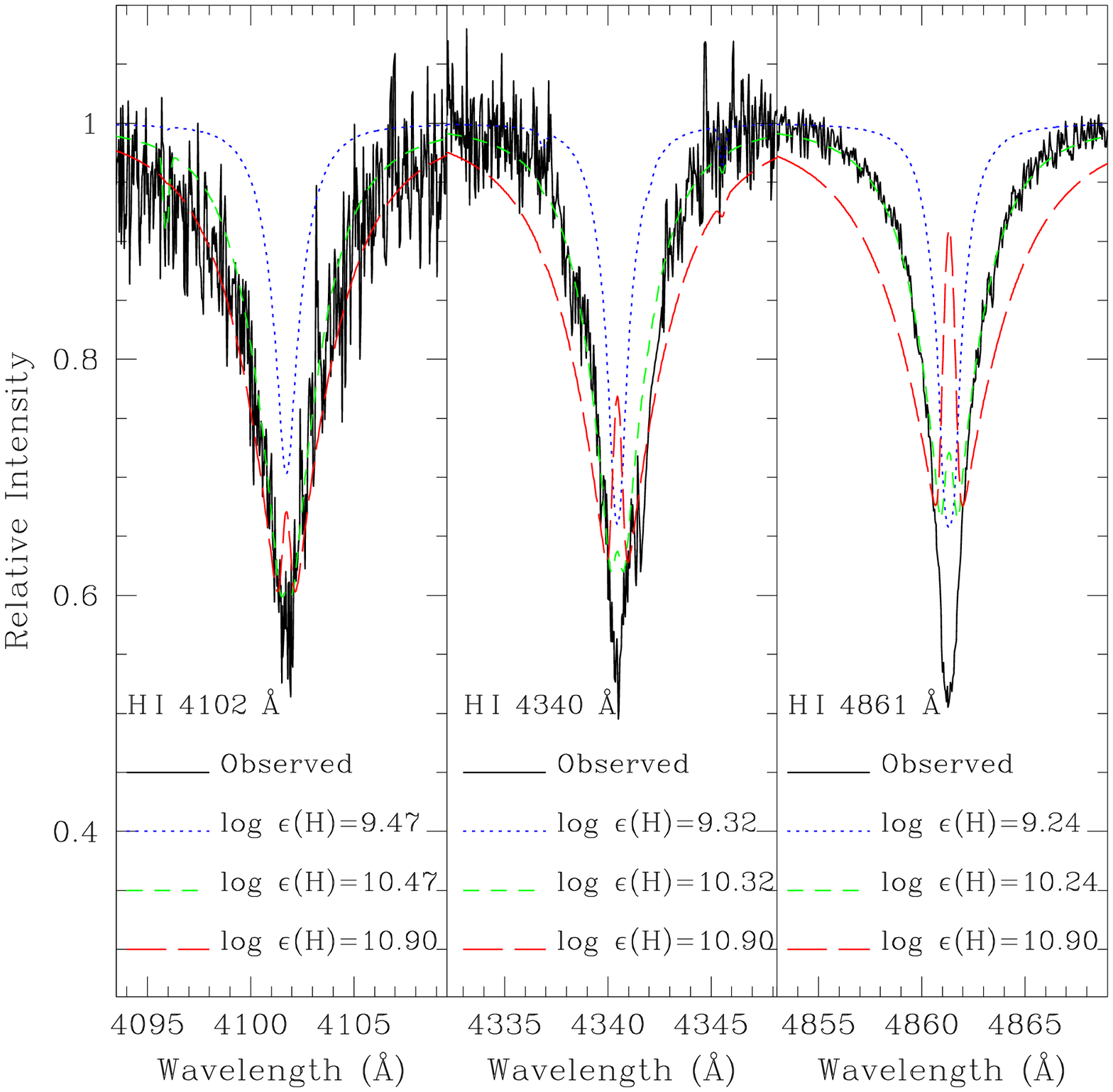}
\caption{The observed spectrum of HD\,144941 and theoretical NLTE H\,{\sc i} line
profiles calculated using the
NLTE model $T_{\rm eff}$=22,000 K and $\log g$ = 3.45, for three different
H abundances $-$ see key on the figure.}
\end{figure}

\begin{figure}
\epsscale{1.00}
\plotone{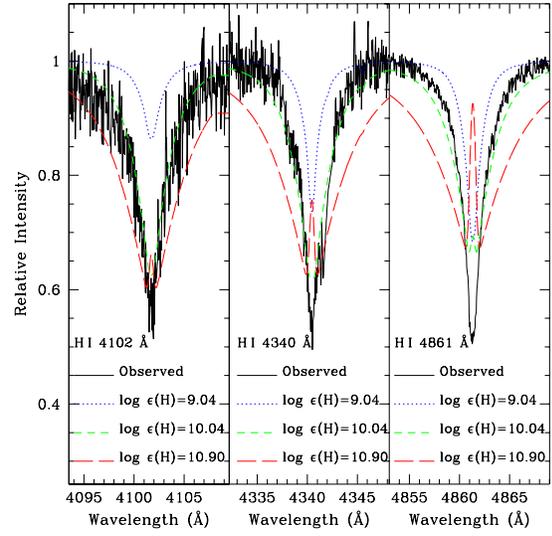}
\caption{The observed spectrum of HD\,144941 and theoretical NLTE H\,{\sc i} line
profiles calculated using a
NLTE model $T_{\rm eff}$=22,000 K and $\log g$ = 4.15, for three different
H abundances $-$ see key on the figure. These model parameters and the H abundance of
10.04 were adopted based on the estimates of \citet{przybilla05}}
\end{figure}

\begin{figure}
\epsscale{1.00}
\plotone{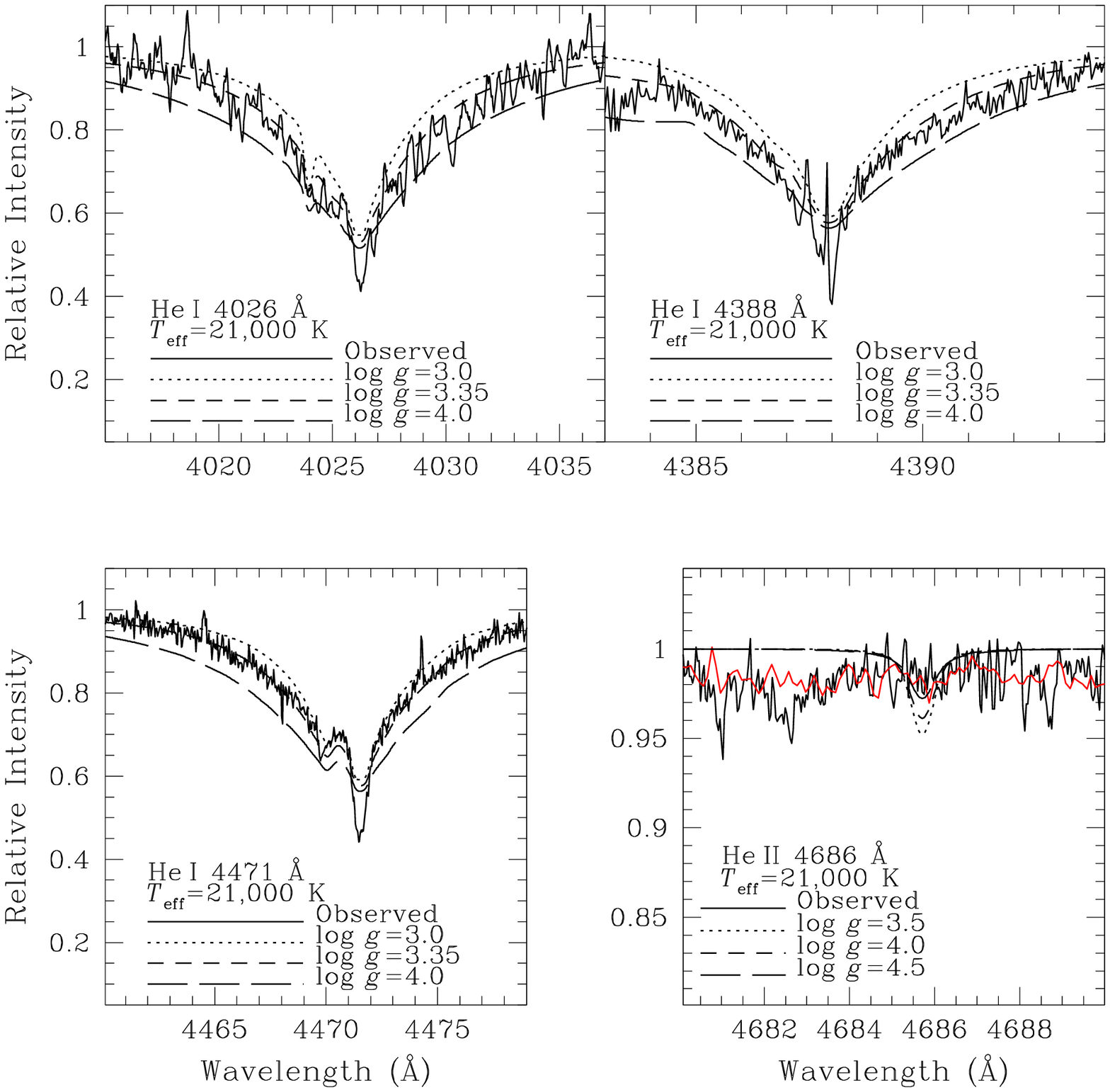}
\caption{The observed spectrum of HD\,144941 and theoretical LTE He\,{\sc i} and
He\,{\sc ii} line profiles calculated using the
LTE model $T_{\rm eff}$=21,000 K, for three different $\log g$ values $-$ see key
on the figure. The AAT spectrum is shown in red.}
\end{figure}

\begin{figure}
\epsscale{1.00}
\plotone{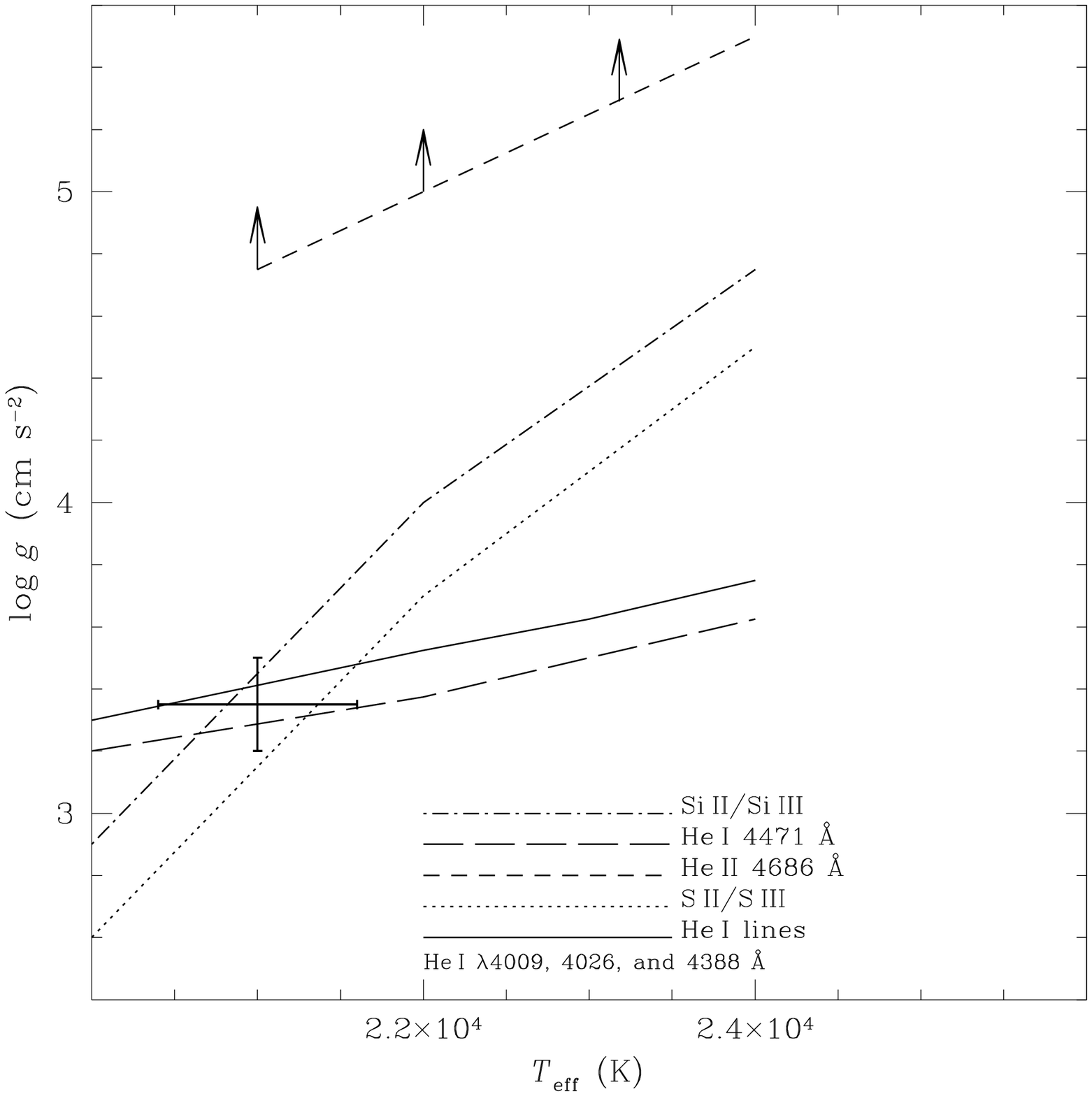}
\caption{The $T_{\rm eff}$ vs $\log g$ plane for
HD\,144941.
Loci satisfying ionization equilibria are plotted $-$ see keys on the figure.
The loci satisfying optical He\,{\sc i} line profiles are shown. 
The limiting locus for He\,{\sc ii} is shown by upward arrows.
The cross shows the adopted LTE model atmosphere parameters.}
\end{figure}

\begin{figure}
\epsscale{1.00}
\plotone{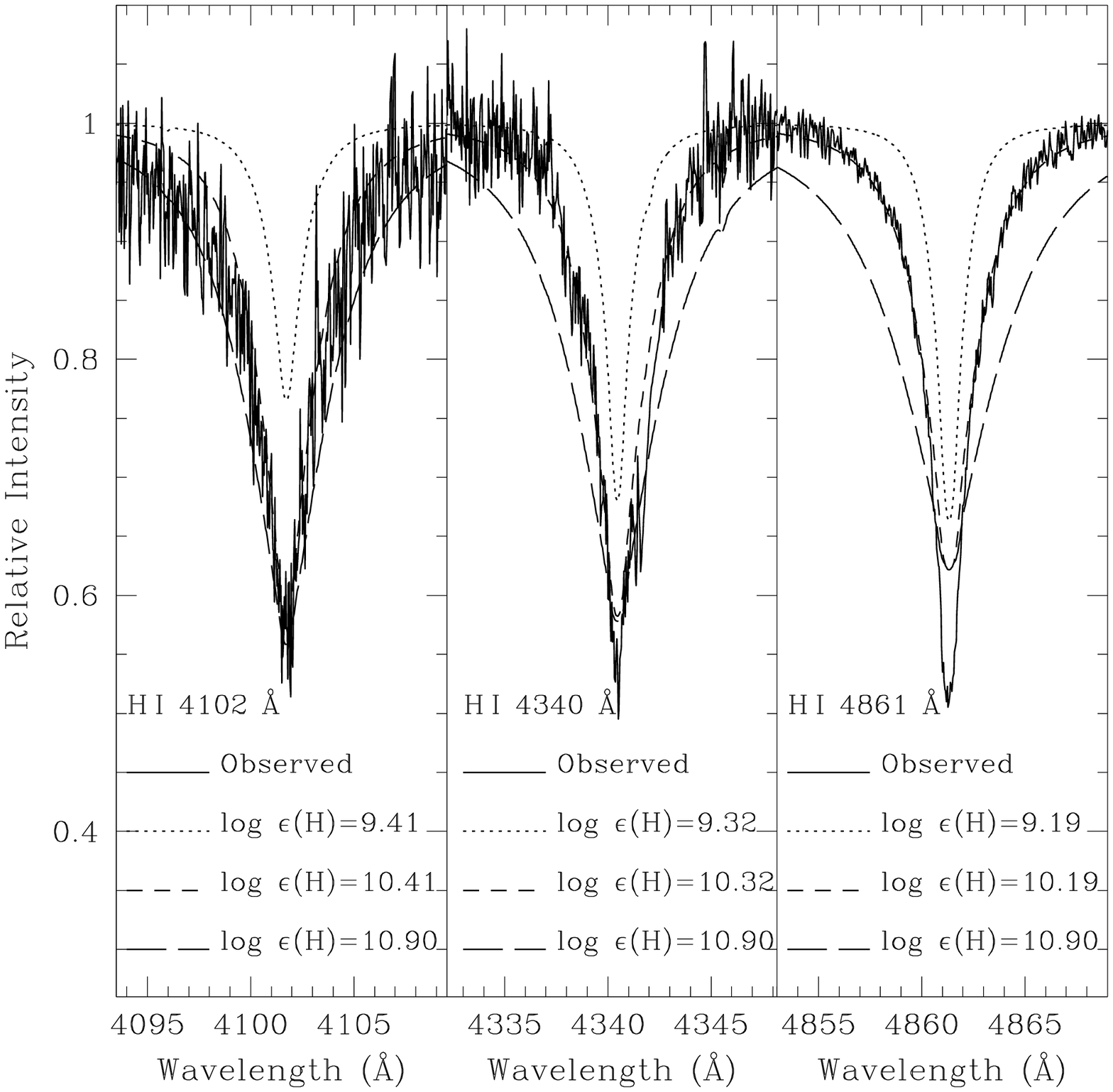}
\caption{The observed spectrum of HD\,144941 and theoretical LTE H\,{\sc i} line
profiles calculated using the LTE model $T_{\rm eff}$=21,000 K and $\log g$ = 3.35, for three different
H abundances $-$ see key on the figure.}
\end{figure}

\begin{figure}
\epsscale{1.00}
\plotone{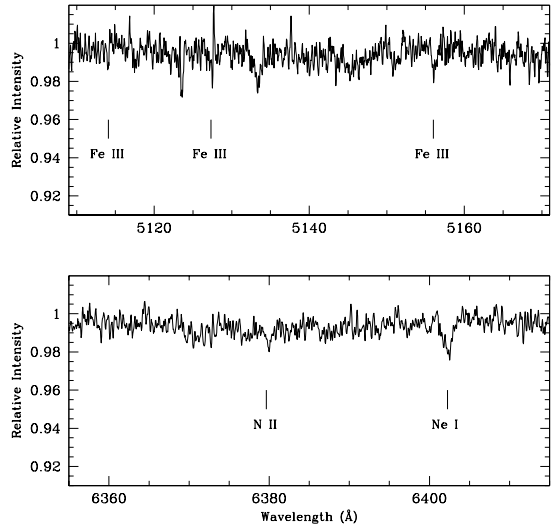}
\caption{Two sample spectral regions are shown for HD\,144941. The positions of the key lines are
marked in particular Ne\,{\sc i} and Fe\,{\sc iii} lines; these lines
are the strongest among the other measured Ne\,{\sc i} and Fe\,{\sc iii} lines, respectively.}
\end{figure}

\begin{deluxetable}{llcccc}
\label{t:equil}
\tablewidth{0pt}
\tablecolumns{7}
\tablecaption{Summary of atmospheric parameters}
\tablehead{
\colhead{} & \colhead{} & \colhead{$T_{\rm eff}$} & \colhead{$\log g$} & \colhead{$\xi$} &
\colhead{$v \sin i$}\\
\colhead{Star} & \colhead{Analyses} & \colhead{(K)} & \colhead{(cgs units)} & \colhead{(km s$^{-1}$)} &
\colhead{(km s$^{-1}$)}}
\startdata
V652\,Her & non-LTE & $25000\pm300$ &  $3.10\pm0.12$ & $13\pm2$ & 10 $-$ 12  \\
V652\,Her & LTE & $25300\pm300$ &  $3.25\pm0.12$ & $13\pm2$ & 10 $-$ 12  \\
HD\,144941 & non-LTE & $22000\pm600$ &  $3.45\pm0.15$ & 10 & 10  \\
HD\,144941 & LTE & $21000\pm600$ &  $3.35\pm0.15$ &  10 & 10  \\
\enddata
\end{deluxetable}

\begin{deluxetable}{lccrcc}
\label{t:lines.uves}
\tabletypesize{\scriptsize}
\tablewidth{0pt}
\tablecolumns{7}
\tablecaption{Measured equivalent widths ($W_{\lambda}$) and NLTE/LTE photospheric line abundances
for V652\,Her.}
\tablehead{
\colhead{} & \colhead{$\chi$} & \colhead{} & \colhead{$W_{\lambda}$} &
\multicolumn{2}{c}{log $\epsilon(\rm X)$}\\
\cline{5-6} \\
\colhead{Line} & \colhead{(eV)} & \colhead{log $gf$} & \colhead{(m\AA)} &
\colhead{NLTE\tablenotemark{a}} & \colhead{LTE\tablenotemark{b}}}
\startdata
H\,{\sc i} $\lambda 4101.734$ & 10.199 & $-$0.753 & Synth & 9.42 &  9.65     \\
H\,{\sc i} $\lambda 4340.462$ & 10.199 & $-$0.447 & Synth & 9.47 &  9.64     \\
H\,{\sc i} $\lambda 4861.323$ & 10.199 & $-$0.020 & Synth & 9.52 &  9.69     \\
        &         &         &         &               &               \\
Mean... & \nodata & \nodata & \nodata &  9.47$\pm$0.05 &  9.66$\pm$0.03 \\
        &         &         &         &               &               \\
C\,{\sc ii} $\lambda 3920.681$* & 16.333 & $-$0.232 & 35  & 7.21 &  6.94     \\
        &         &         &         &               &               \\
C\,{\sc ii} $\lambda 4267.001$* & 18.046 & $+$0.563 &     &      &           \\
C\,{\sc ii} $\lambda 4267.183$* & 18.046 & $+$0.716 &     &      &           \\
C\,{\sc ii} $\lambda 4267.261$* & 18.046 & $-$0.584 & 125 & 7.29 &  6.94     \\
        &         &         &         &               &               \\
C\,{\sc ii} $\lambda 6578.050$* & 14.449 & $-$0.026 & 69  & 7.04 &  6.86     \\
C\,{\sc ii} $\lambda 6582.880$* & 14.449 & $-$0.327 & 45  & 7.08 &  6.90     \\
        &         &         &         &               &               \\
Mean... & \nodata & \nodata & \nodata & 7.16$\pm$0.12 & 6.91$\pm$0.04 \\
        &         &         &         &               &               \\
C\,{\sc iii} $\lambda 4647.418$* & 29.535 &$+$0.070 & 25  & 6.86 & 6.88 \\
C\,{\sc iii} $\lambda 4650.246$* & 29.535 &$-$0.151 & 22  & 6.92 & 7.02 \\
        &         &         &         &               &               \\
Mean... & \nodata & \nodata & \nodata & 6.90$\pm$0.04 & 6.95$\pm$0.10 \\
        &         &         &         &               &               \\
N\,{\sc ii} $\lambda 3955.851$* & 18.466 & $-$0.849 & 139 &        & 8.56 \\
N\,{\sc ii} $\lambda 3994.997$* & 18.497 & $+$0.163 & 308 &  8.75  & 8.99 \\
N\,{\sc ii} $\lambda 4227.736$* & 21.600 & $-$0.061 & 153 &        & 8.64 \\
N\,{\sc ii} $\lambda 4459.937$* & 20.646 & $-$1.476 &  28 &  8.77  & 8.63 \\
N\,{\sc ii} $\lambda 4507.560$* & 20.666 & $-$0.817 &  74 &        & 8.54 \\
N\,{\sc ii} $\lambda 4564.760$* & 20.409 & $-$1.589 &  30 &        & 8.73 \\
N\,{\sc ii} $\lambda 4601.478$* & 18.466 & $-$0.452 & 220 &  8.73  & 8.87 \\
N\,{\sc ii} $\lambda 4607.153$* & 18.462 & $-$0.522 & 206 &  8.73  & 8.83 \\
N\,{\sc ii} $\lambda 4613.868$* & 18.466 & $-$0.665 & 194 &  8.79  & 8.86 \\
N\,{\sc ii} $\lambda 4621.393$* & 18.466 & $-$0.538 & 215 &  8.79  & 8.91 \\
N\,{\sc ii} $\lambda 4630.539$* & 18.483 & $+$0.080 & 316 &  8.78  &      \\
N\,{\sc ii} $\lambda 4643.086$* & 18.483 & $-$0.371 & 222 &  8.55  & 8.82 \\
N\,{\sc ii} $\lambda 4654.531$* & 18.497 & $-$1.506 &  85 &        & 8.83 \\
N\,{\sc ii} $\lambda 4667.208$* & 18.497 & $-$1.646 &  65 &        & 8.79 \\
N\,{\sc ii} $\lambda 4674.908$* & 18.497 & $-$1.553 &  80 &        & 8.84 \\
N\,{\sc ii} $\lambda 4718.377$ & 27.746 & $-$0.042 &  33 &        & 8.91 \\
N\,{\sc ii} $\lambda 4774.244$* & 20.646 & $-$1.280 &  51 &        & 8.78 \\
N\,{\sc ii} $\lambda 4779.722$* & 20.646 & $-$0.587 & 106 &        & 8.60 \\
N\,{\sc ii} $\lambda 4781.190$* & 20.654 & $-$1.337 &  40 &        & 8.70 \\
N\,{\sc ii} $\lambda 4788.138$* & 20.654 & $-$0.363 & 120 &        & 8.49 \\
N\,{\sc ii} $\lambda 4803.287$* & 20.666 & $-$0.113 & 160 &        & 8.57 \\
N\,{\sc ii} $\lambda 4810.299$* & 20.666 & $-$1.084 &  54 &        & 8.63 \\
N\,{\sc ii} $\lambda 4987.376$* & 20.940 & $-$0.584 &  87 &        & 8.53 \\
N\,{\sc ii} $\lambda 4991.243$* & 25.491 & $-$0.180 &  50 &        & 8.81 \\
N\,{\sc ii} $\lambda 4997.224$* & 25.491 & $-$0.657 &  28 &        & 8.96 \\
N\,{\sc ii} $\lambda 5002.703$* & 18.462 & $-$1.022 & 126 &        & 8.65 \\
N\,{\sc ii} $\lambda 5023.053$* & 25.507 & $-$0.165 &  60 &        & 8.91 \\
N\,{\sc ii} $\lambda 5045.099$* & 18.483 & $-$0.407 & 245 &  8.66  & 8.83 \\
N\,{\sc ii} $\lambda 5073.592$* & 18.497 & $-$1.550 &  85 &        & 8.91 \\
        &         &         &         &               &               \\
N\,{\sc ii} $\lambda 5179.344$ & 27.980 & $+$0.497 &     &        &      \\
N\,{\sc ii} $\lambda 5179.521$ & 27.746 & $+$0.675 &  95 &        & 8.71 \\
        &         &         &         &               &               \\
N\,{\sc ii} $\lambda 5183.200$ & 27.980 & $-$0.090 &  26 &        & 8.94 \\
N\,{\sc ii} $\lambda 5184.961$ & 27.739 & $-$0.044 &  26 &        & 8.84 \\
N\,{\sc ii} $\lambda 5452.070$* & 21.148 & $-$0.881 &  60 &  8.74  & 8.66 \\
N\,{\sc ii} $\lambda 5454.215$* & 21.153 & $-$0.782 &  97 &  8.88  & 8.91 \\
N\,{\sc ii} $\lambda 5478.086$* & 21.153 & $-$0.930 &  40 &        & 8.47 \\
N\,{\sc ii} $\lambda 5480.050$* & 21.160 & $-$0.711 &  56 &        & 8.45 \\
N\,{\sc ii} $\lambda 5495.655$* & 21.160 & $-$0.220 & 106 &        & 8.43 \\
N\,{\sc ii} $\lambda 5526.234$* & 25.491 & $-$0.312 &  45 &  8.58  & 8.93 \\
N\,{\sc ii} $\lambda 5530.242$* & 25.498 & $+$0.113 &  81 &  8.48  & 8.90 \\
        &         &         &         &               &               \\
N\,{\sc ii} $\lambda 5535.347$* & 25.507 & $+$0.398 &     &        &      \\
N\,{\sc ii} $\lambda 5535.383$* & 25.491 & $-$0.204 & 135 &  8.43  & 8.98 \\
        &         &         &         &               &               \\
N\,{\sc ii} $\lambda 5540.061$* & 25.491 & $-$0.557 &  32 &  8.65  & 8.98 \\
N\,{\sc ii} $\lambda 5543.471$* & 25.498 & $-$0.092 &  68 &  8.59  & 8.98 \\
N\,{\sc ii} $\lambda 5551.922$* & 25.507 & $-$0.189 &  63 &  8.64  & 9.03 \\
N\,{\sc ii} $\lambda 5676.020$* & 18.462 & $-$0.367 & 236 &  8.77  & 9.06 \\
N\,{\sc ii} $\lambda 5710.770$* & 18.483 & $-$0.518 & 199 &  8.87  & 8.88 \\
N\,{\sc ii} $\lambda 5730.660$* & 18.483 & $-$1.703 &  51 &        & 8.77 \\
N\,{\sc ii} $\lambda 5747.300$* & 18.497 & $-$1.091 & 116 &        & 8.77 \\
N\,{\sc ii} $\lambda 5767.450$* & 18.497 & $-$1.447 &  95 &        & 8.95 \\
N\,{\sc ii} $\lambda 6136.890$* & 23.132 & $-$1.124 &  23 &        & 8.89 \\
N\,{\sc ii} $\lambda 6150.750$* & 23.125 & $-$1.086 &  24 &        & 8.87 \\
N\,{\sc ii} $\lambda 6167.750$* & 23.142 & $+$0.025 & 122 &        & 8.86 \\
N\,{\sc ii} $\lambda 6170.160$* & 23.125 & $-$0.311 &  73 &        & 8.75 \\
N\,{\sc ii} $\lambda 6173.310$* & 23.132 & $-$0.126 & 100 &        & 8.82 \\
N\,{\sc ii} $\lambda 6340.580$* & 23.246 & $-$0.192 &  79 &        & 8.73 \\
N\,{\sc ii} $\lambda 6346.860$* & 23.239 & $-$0.901 &  37 &        & 8.96 \\
N\,{\sc ii} $\lambda 6379.620$* & 18.466 & $-$1.191 & 103 &  8.60  & 8.81 \\
N\,{\sc ii} $\lambda 6482.050$* & 18.497 & $-$0.311 & 229 &  8.82  & 9.04 \\
N\,{\sc ii} $\lambda 6504.610$* & 23.246 & $-$0.626 &  55 &        & 8.93 \\
        &         &         &         &               &               \\
Mean... & \nodata & \nodata & \nodata & 8.70$\pm$0.12 & 8.80$\pm$0.16 \\
        &         &         &         &               &               \\
N\,{\sc iii} $\lambda 4514.850$* & 35.671 & $+$0.221 &  46 &        & 9.13 \\
N\,{\sc iii} $\lambda 4634.130$* & 30.459 & $-$0.086 &  78 &  8.52  & 8.81 \\
N\,{\sc iii} $\lambda 4640.640$* & 30.463 & $+$0.168 & 100 &  8.41  & 8.81 \\
        &         &         &         &               &               \\
Mean... & \nodata & \nodata & \nodata & 8.50$\pm$0.10 & 8.92$\pm$0.18 \\
        &         &         &         &               &               \\
O\,{\sc ii} $\lambda 4345.560$* & 22.979 & $-$0.346  &  62 & 7.64  & 7.67 \\
O\,{\sc ii} $\lambda 4366.895$* & 22.999 & $-$0.348  &  60 & 7.60  & 7.64 \\
O\,{\sc ii} $\lambda 4414.899$* & 23.441 & $+$0.172  & 104 & 7.53  & 7.64 \\
O\,{\sc ii} $\lambda 4416.975$* & 23.419 & $-$0.077  & 107 & 7.78  & 7.91 \\
O\,{\sc ii} $\lambda 4452.378$* & 23.442 & $-$0.788  &  23 & 7.56  & 7.65 \\
O\,{\sc ii} $\lambda 4590.974$* & 25.661 & $+$0.350  &  70 & 7.79  & 7.68 \\
O\,{\sc ii} $\lambda 4596.177$* & 25.661 & $+$0.200  &  52 & 7.72  & 7.63 \\
O\,{\sc ii} $\lambda 4638.856$* & 22.966 & $-$0.332  &  64 & 7.64  & 7.69 \\
O\,{\sc ii} $\lambda 4649.135$* & 22.999 & $+$0.308  & 140 & 7.64  & 7.74 \\
O\,{\sc ii} $\lambda 4650.838$* & 22.966 & $-$0.362  &  38 & 7.35  & 7.40 \\
O\,{\sc ii} $\lambda 4661.632$* & 22.979 & $-$0.278  &  62 & 7.57  & 7.62 \\
O\,{\sc ii} $\lambda 4676.235$* & 22.999 & $-$0.394  &  42 & 7.44  & 7.50 \\
        &         &         &         &               &               \\
O\,{\sc ii} $\lambda 4699.011$* & 28.510 & $+$0.418  &     &       &      \\
O\,{\sc ii} $\lambda 4699.218$* & 26.225 & $+$0.270  &  36 & 7.37  & 7.30 \\
        &         &         &         &               &               \\
O\,{\sc ii} $\lambda 4705.346$* & 26.249 & $+$0.477  &  44 & 7.43  & 7.40 \\
O\,{\sc ii} $\lambda 5206.651$* & 26.561 & $-$0.266  &  15 & 7.73  & 7.69 \\
        &         &         &         &               &               \\
Mean... & \nodata & \nodata & \nodata & 7.59$\pm$0.14 & 7.61$\pm$0.15 \\
        &         &         &         &               &               \\
Ne\,{\sc i} $\lambda 5852.488$* & 16.848 & $-$0.490  & 20  &       & 8.51 \\
Ne\,{\sc i} $\lambda 6143.063$* & 16.619 & $-$0.100  & 53  & 8.13  & 8.62 \\
Ne\,{\sc i} $\lambda 6163.594$* & 16.715 & $-$0.620  & 24  & 7.98  & 8.72 \\
Ne\,{\sc i} $\lambda 6266.495$* & 16.715 & $-$0.370  & 25  & 7.75  & 8.50 \\
Ne\,{\sc i} $\lambda 6334.428$* & 16.619 & $-$0.320  & 40  & 8.23  & 8.69 \\
Ne\,{\sc i} $\lambda 6382.991$* & 16.671 & $-$0.240  & 40  & 8.13  & 8.62 \\
Ne\,{\sc i} $\lambda 6402.246$* & 16.619 & $+$0.330  & 104 & 8.09  & 8.72 \\
Ne\,{\sc i} $\lambda 6506.528$* & 16.671 & $-$0.030  & 52  & 8.07  & 8.58 \\
Ne\,{\sc i} $\lambda 6598.953$* & 16.848 & $-$0.360  & 22  & 7.97  & 8.48 \\
Ne\,{\sc i} $\lambda 7032.413$* & 16.619 & $-$0.260  & 40  & 8.16  & 8.67 \\
        &         &         &         &               &               \\
Mean... & \nodata & \nodata & \nodata & 8.06$\pm$0.14 & 8.61$\pm$0.09 \\
        &         &         &         &               &               \\
Mg\,{\sc ii} $\lambda 4481.126$* &  8.864 & $+$0.749  &     &       &      \\
Mg\,{\sc ii} $\lambda 4481.150$* &  8.864 & $-$0.553  &     &       &      \\
Mg\,{\sc ii} $\lambda 4481.325$* &  8.864 & $+$0.594  & 211 & 7.09  & 7.48 \\
        &         &         &         &               &               \\
Al\,{\sc iii} $\lambda 4149.913$ & 20.555 & $+$0.620  &     &       &      \\
Al\,{\sc iii} $\lambda 4149.968$ & 20.555 & $-$0.680  &     &       &      \\
Al\,{\sc iii} $\lambda 4150.173$ & 20.555 & $+$0.470  & 103 &       & 6.40 \\
        &         &         &         &               &               \\
Al\,{\sc iii} $\lambda 4479.885$ & 20.781 & $+$0.900\tablenotemark{c}  &     &       &       \\
Al\,{\sc iii} $\lambda 4479.971$ & 20.781 & $+$1.020\tablenotemark{c}  &     &       &       \\
Al\,{\sc iii} $\lambda 4480.009$ & 20.781 & $-$0.530\tablenotemark{c}  & 75  &       & 5.93 \\
        &         &         &         &               &               \\
Al\,{\sc iii} $\lambda 4512.565$ & 17.808 & $+$0.410  & 89  &       & 6.24 \\
        &         &         &         &               &               \\
Al\,{\sc iii} $\lambda 4528.945$ & 17.818 & $-$0.290  &     &       &      \\
Al\,{\sc iii} $\lambda 4529.189$ & 17.818 & $+$0.660  & 146 &       & 6.34 \\
        &         &         &         &               &               \\
Al\,{\sc iii} $\lambda 5696.604$ & 15.642 & $+$0.230  & 169 &       & 6.67 \\
Al\,{\sc iii} $\lambda 5722.730$ & 15.642 & $-$0.070  & 126 &       & 6.60 \\
        &         &         &         &               &               \\
Mean... & \nodata & \nodata & \nodata &               & 6.36$\pm$0.27 \\
        &         &         &         &               &               \\
Si\,{\sc ii} $\lambda 4128.054$ &  9.837 & $+$0.359  &  30 &       & 7.19 \\
        &         &         &         &               &               \\
Si\,{\sc ii} $\lambda 4130.872$ &  9.839 & $-$0.783  &     &       &      \\
Si\,{\sc ii} $\lambda 4130.894$ &  9.839 & $+$0.552  &  34 &       & 7.04 \\
        &         &         &         &               &               \\
Si\,{\sc ii} $\lambda 5041.024$ & 10.066 & $+$0.029  &  35 &       & 7.67 \\
Si\,{\sc ii} $\lambda 5055.984$ & 10.074 & $+$0.523  &  44 &       & 7.30 \\
        &         &         &         &               &               \\
Mean... & \nodata & \nodata & \nodata &               & 7.30$\pm$0.27 \\
        &         &         &         &               &               \\
Si\,{\sc iii} $\lambda 3796.124$* & 21.730 & $+$0.407  &     &       &      \\
Si\,{\sc iii} $\lambda 3796.203$* & 21.730 & $-$0.703  & 167 & 7.18  & 7.04 \\
        &         &         &         &               &               \\
Si\,{\sc iii} $\lambda 3806.526$* & 21.739 & $+$0.679  &     &       &      \\
Si\,{\sc iii} $\lambda 3806.700$* & 21.739 & $-$0.071  & 307 & 7.60  & 7.69 \\
        &         &         &         &               &               \\
Si\,{\sc iii} $\lambda 4567.840$* & 19.016 & $+$0.068  & 285 & 7.40  & 7.85 \\
Si\,{\sc iii} $\lambda 4574.757$* & 19.016 & $-$0.409  & 204 & 7.34  & 7.57 \\
Si\,{\sc iii} $\lambda 4716.654$* & 25.334 & $+$0.491  &  83 & 7.46  & 7.15 \\
Si\,{\sc iii} $\lambda 4813.333$* & 25.979 & $+$0.708  &  85 & 7.04  & 7.11 \\
        &         &         &         &               &               \\
Si\,{\sc iii} $\lambda 4819.712$* & 25.982 & $+$0.937  &     &       &      \\
Si\,{\sc iii} $\lambda 4819.814$* & 25.982 & $-$0.354  & 116 & 7.01  & 7.24 \\
        &         &         &         &               &               \\
Si\,{\sc iii} $\lambda 4828.951$* & 25.987 & $+$0.937  &     &       &      \\
Si\,{\sc iii} $\lambda 4829.111$* & 25.980 & $-$0.354  & 120 & 7.03  & 7.16 \\
        &         &         &         &               &               \\
Si\,{\sc iii} $\lambda 5739.734$* & 19.722 & $-$0.096  & 226 & 7.36  & 7.76 \\
        &         &         &         &               &               \\
Mean... & \nodata & \nodata & \nodata & 7.27$\pm$0.21 & 7.40$\pm$0.32 \\
        &         &         &         &               &               \\
Si\,{\sc iv} $\lambda 4088.862$* & 24.050 & $+$0.194  & 185 & 7.63  & 7.75 \\
Si\,{\sc iv} $\lambda 4116.104$* & 24.050 & $-$0.110  & 120 & 7.48  & 7.44 \\
        &         &         &         &               &               \\
Mean... & \nodata & \nodata & \nodata & 7.56$\pm$0.11 & 7.60$\pm$0.22 \\
        &         &         &         &               &               \\
P\,{\sc iii} $\lambda 4222.198$ & 14.610 & $+$0.210  &  80 &       & 5.42 \\
P\,{\sc iii} $\lambda 4246.720$ & 14.610 & $-$0.120  &  65 &       & 5.61 \\
        &         &         &         &               &               \\
Mean... & \nodata & \nodata & \nodata &               & 5.52$\pm$0.13 \\
        &         &         &         &               &               \\
S\,{\sc ii} $\lambda 5032.434$ & 13.672 & $+$0.188  &  40 &       & 7.35 \\
S\,{\sc ii} $\lambda 5103.332$ & 13.672 & $-$0.457  &  25 &       & 7.76 \\
S\,{\sc ii} $\lambda 5212.620$ & 15.068 & $+$0.316  &  25 & 7.21  & 7.33 \\
S\,{\sc ii} $\lambda 5320.723$ & 15.068 & $+$0.431  &  27 & 7.12  & 7.26 \\
S\,{\sc ii} $\lambda 5428.655$* & 13.584 & $-$0.177  &  20 & 7.57  & 7.37 \\
S\,{\sc ii} $\lambda 5432.797$* & 13.617 & $+$0.205  &  35 & 7.46  & 7.27 \\
S\,{\sc ii} $\lambda 5509.705$* & 13.617 & $-$0.175  &  20 & 7.56  & 7.37 \\
S\,{\sc ii} $\lambda 5564.958$ & 13.672 & $-$0.336  &  33 &       & 7.80 \\
S\,{\sc ii} $\lambda 5606.151$ & 13.733 & $+$0.124  &  25 &       & 7.22 \\
        &         &         &         &               &               \\
S\,{\sc ii} $\lambda 5639.977$ & 14.067 & $+$0.258  &     &       &      \\
S\,{\sc ii} $\lambda 5640.346$ & 13.701 & $-$0.036  &  45 &       & 7.23 \\
        &         &         &         &               &               \\
S\,{\sc ii} $\lambda 5660.001$ & 13.677 & $-$0.222  &  17 &       & 7.37 \\
        &         &         &         &               &               \\
Mean... & \nodata & \nodata & \nodata & 7.38$\pm$0.21 & 7.39$\pm$0.20 \\
        &         &         &         &               &               \\
S\,{\sc iii} $\lambda 4253.589$* & 18.244 & $+$0.107  & 174 & 7.46  & 7.18 \\
S\,{\sc iii} $\lambda 4284.979$* & 18.193 & $-$0.233  & 103 & 7.28  & 6.90 \\
S\,{\sc iii} $\lambda 4332.692$* & 18.188 & $-$0.564  &  66 & 7.26  & 6.88 \\
S\,{\sc iii} $\lambda 4354.566$* & 18.311 & $-$0.959  &  53 & 7.49  & 7.15 \\
S\,{\sc iii} $\lambda 4361.527$* & 18.244 & $-$0.606  &  68 & 7.32  & 6.95 \\
S\,{\sc iii} $\lambda 4364.747$* & 18.318 & $-$0.805  &  34 & 7.07  & 6.74 \\
S\,{\sc iii} $\lambda 4499.245$* & 18.294 & $-$1.640  &  16 & 7.52  & 7.20 \\
        &         &         &         &               &               \\
Mean... & \nodata & \nodata & \nodata & 7.34$\pm$0.16 & 7.00$\pm$0.18 \\
        &         &         &         &               &               \\
Ar\,{\sc ii} $\lambda 4806.021$ & 16.644 & $+$0.210  &  32  &       & 6.91 \\
        &         &         &         &               &               \\
Fe\,{\sc iii} $\lambda 4137.764$ & 20.613 & $+$0.630\tablenotemark{c}  &  42  & 6.74  & 7.05 \\
Fe\,{\sc iii} $\lambda 4164.731$ & 20.634 & $+$0.923\tablenotemark{c}  &  83  & 6.91  & 7.20 \\
        &         &         &         &               &               \\
Fe\,{\sc iii} $\lambda 4296.851$ & 22.860 & $+$0.418\tablenotemark{c}  &      &       &      \\
Fe\,{\sc iii} $\lambda 4296.851$ & 22.860 & $+$0.879\tablenotemark{c}  &  30  & 6.71  & 7.02 \\
        &         &         &         &               &               \\
Fe\,{\sc iii} $\lambda 4310.355$ & 22.869 & $+$0.189\tablenotemark{c}  &      &       &      \\
Fe\,{\sc iii} $\lambda 4310.355$ & 22.869 & $+$1.156\tablenotemark{c}  &  45  & 6.76  & 7.05 \\
        &         &         &         &               &               \\
Fe\,{\sc iii} $\lambda 4395.755$ &  8.256 & $-$2.595\tablenotemark{c}  &  39  & 7.04  & 7.34 \\
Fe\,{\sc iii} $\lambda 4419.596$ &  8.241 & $-$2.218\tablenotemark{c}  &  81  & 7.15  & 7.38 \\
Fe\,{\sc iii} $\lambda 4431.019$ &  8.248 & $-$2.572\tablenotemark{c}  &  38  & 7.02  & 7.30 \\
Fe\,{\sc iii} $\lambda 5063.421$ &  8.648 & $-$2.950\tablenotemark{c}  &  18  & 7.14  & 7.42 \\
Fe\,{\sc iii} $\lambda 5086.701$ &  8.659 & $-$2.590\tablenotemark{c}  &  40  & 7.17  & 7.46 \\
Fe\,{\sc iii} $\lambda 5127.387$ &  8.659 & $-$2.218\tablenotemark{c}  & 100  &       & 7.66 \\
Fe\,{\sc iii} $\lambda 5156.111$ &  8.641 & $-$2.018\tablenotemark{c}  & 100  & 7.31  & 7.46 \\
Fe\,{\sc iii} $\lambda 5235.658$ & 18.266 & $-$0.107\tablenotemark{c}  &  33  &       & 7.18 \\
Fe\,{\sc iii} $\lambda 5243.306$ & 18.270 & $+$0.405\tablenotemark{c}  &  71  &       & 7.13 \\
Fe\,{\sc iii} $\lambda 5276.476$ & 18.264 & $-$0.001\tablenotemark{c}  &  36  &       & 7.13 \\
Fe\,{\sc iii} $\lambda 5282.297$ & 18.266 & $+$0.108\tablenotemark{c}  &  43  &       & 7.12 \\
Fe\,{\sc iii} $\lambda 5299.926$ & 18.261 & $-$0.166\tablenotemark{c}  &  31  &       & 7.21 \\
Fe\,{\sc iii} $\lambda 5302.602$ & 18.262 & $-$0.120\tablenotemark{c}  &  31  &       & 7.17 \\
Fe\,{\sc iii} $\lambda 5460.799$ & 14.178 & $-$1.519\tablenotemark{c}  &  30  & 6.88  & 7.59 \\
Fe\,{\sc iii} $\lambda 5485.517$ & 14.176 & $-$1.469\tablenotemark{c}  &  30  & 6.83  & 7.54 \\
Fe\,{\sc iii} $\lambda 5573.424$ & 14.175 & $-$1.390\tablenotemark{c}  &  42  & 6.92  & 7.64 \\
Fe\,{\sc iii} $\lambda 5833.938$ & 18.509 & $+$0.616\tablenotemark{c}  &  66  &       & 6.96 \\
Fe\,{\sc iii} $\lambda 5891.904$ & 18.509 & $+$0.498\tablenotemark{c}  &  54  &       & 6.96 \\
Fe\,{\sc iii} $\lambda 5929.685$ & 18.509 & $+$0.351\tablenotemark{c}  &  48  &       & 7.04 \\
        &         &         &         &               &               \\
Mean... & \nodata & \nodata & \nodata & 7.05$\pm$0.16 & 7.30$\pm$0.22 \\
\enddata
\tablenotetext{a}{($T_{\rm eff}$, $\log g$, $\xi$)=(25000, 3.10, 13.0)}
\tablenotetext{b}{($T_{\rm eff}$, $\log g$, $\xi$)=(25300, 3.25, 13.0)}
\tablenotetext{c}{Kurucz $gf$-value}
\tablenotetext{*}{The lines covered by the adopted model atom, including the
extended model atom with more levels, for the ions providing the ionization balance and some other ions}
\end{deluxetable}

\begin{deluxetable}{lccrcc}
\label{t:lines.uvesh}
\tabletypesize{\scriptsize}
\tablewidth{0pt}
\tablecolumns{7}
\tablecaption{Measured equivalent widths ($W_{\lambda}$) and NLTE/LTE photospheric line abundances
for HD\,144941.}
\tablehead{
\colhead{} & \colhead{$\chi$} & \colhead{} & \colhead{$W_{\lambda}$} &
\multicolumn{2}{c}{log $\epsilon(\rm X)$}\\
\cline{5-6} \\
\colhead{Line} & \colhead{(eV)} & \colhead{log $gf$} & \colhead{(m\AA)} & \colhead{NLTE\tablenotemark{a}} &
\colhead{LTE\tablenotemark{b}}}
\startdata
H\,{\sc i} $\lambda 4101.734$ & 10.199 & $-$0.753 & Synth &10.47 & 10.41     \\
H\,{\sc i} $\lambda 4340.462$ & 10.199 & $-$0.447 & Synth &10.32 & 10.32     \\
H\,{\sc i} $\lambda 4861.323$ & 10.199 & $-$0.020 & Synth &10.24 & 10.19     \\
        &         &         &         &               &               \\
Mean... & \nodata & \nodata & \nodata & 10.37$\pm$0.09 & 10.35$\pm$0.05 \\
        &         &         &         &               &               \\
C\,{\sc ii} $\lambda 6578.050$* & 14.449 & $-$0.026 & 88  & 6.88 &  6.64     \\
C\,{\sc ii} $\lambda 6582.880$* & 14.449 & $-$0.327 & 52  & 6.84 &  6.58     \\
        &         &         &         &               &               \\
Mean... & \nodata & \nodata & \nodata & 6.86$\pm$0.03 & 6.61$\pm$0.04 \\
        &         &         &         &               &               \\
N\,{\sc ii} $\lambda 3994.997$* & 18.497 & $+$0.163 &  74 &  6.68  & 6.98 \\
N\,{\sc ii} $\lambda 4447.030$* & 20.409 & $+$0.221 &  22 &  6.50  & 6.72 \\
N\,{\sc ii} $\lambda 4601.478$* & 18.466 & $-$0.452 &  16 &  6.49  & 6.75 \\
N\,{\sc ii} $\lambda 4607.153$* & 18.462 & $-$0.522 &  15 &  6.54  & 6.80 \\
N\,{\sc ii} $\lambda 4621.393$* & 18.466 & $-$0.538 &  15 &  6.54  & 6.81 \\
N\,{\sc ii} $\lambda 4630.539$* & 18.483 & $+$0.080 &  35 &  6.36  & 6.64 \\
        &         &         &         &               &               \\
N\,{\sc ii} $\lambda 4994.360$* & 25.498 & $-$0.164 &     &        &      \\
N\,{\sc ii} $\lambda 4994.370$* & 20.940 & $-$0.098 &   8 &  6.55  & 6.75 \\
        &         &         &         &               &               \\
N\,{\sc ii} $\lambda 5001.474$* & 20.654 & $+$0.435 &  24 &  6.47  & 6.69 \\
N\,{\sc ii} $\lambda 5025.659$* & 20.666 & $-$0.558 &   8 &  6.94  & 7.15 \\
N\,{\sc ii} $\lambda 5045.099$* & 18.483 & $-$0.407 &  16 &  6.49  & 6.76 \\
N\,{\sc ii} $\lambda 5666.630$* & 18.466 & $-$0.080 &  12 &  6.10  & 6.37 \\
N\,{\sc ii} $\lambda 5676.020$* & 18.462 & $-$0.367 &   8 &  6.19  & 6.47 \\
N\,{\sc ii} $\lambda 5679.560$* & 18.483 & $+$0.250 &  24 &  6.11  & 6.39 \\
N\,{\sc ii} $\lambda 6482.050$* & 18.497 & $-$0.311 &   7 &  6.17  & 6.45 \\
        &         &         &         &               &               \\
Mean... & \nodata & \nodata & \nodata & 6.44$\pm$0.23 & 6.70$\pm$0.22 \\
        &         &         &         &               &               \\
O\,{\sc ii} $\lambda 4414.899$* & 23.441 & $+$0.172  &  19 & 6.94  & 6.87 \\
O\,{\sc ii} $\lambda 4416.975$* & 23.419 & $-$0.077  &  16 & 7.10  & 7.02 \\
O\,{\sc ii} $\lambda 4590.974$* & 25.661 & $+$0.350  &  14 & 7.24  & 7.10 \\
O\,{\sc ii} $\lambda 4596.177$* & 25.661 & $+$0.200  &  11 & 7.26  & 7.12 \\
O\,{\sc ii} $\lambda 4638.856$* & 22.966 & $-$0.332  &   8 & 6.94  & 6.86 \\
O\,{\sc ii} $\lambda 4649.135$* & 22.999 & $+$0.308  &  37 & 7.15  & 7.08 \\
O\,{\sc ii} $\lambda 4650.838$* & 22.966 & $-$0.362  &   7 & 6.91  & 6.83 \\
        &         &         &         &               &               \\
Mean... & \nodata & \nodata & \nodata & 7.08$\pm$0.15 & 6.98$\pm$0.13 \\
        &         &         &         &               &               \\
Ne\,{\sc i} $\lambda 6143.063$* & 16.619 & $-$0.100  & 12  & 7.26  & 7.49 \\
Ne\,{\sc i} $\lambda 6402.246$* & 16.619 & $+$0.330  & 28  & 7.26  & 7.50 \\
Ne\,{\sc i} $\lambda 6506.528$* & 16.671 & $-$0.030  & 10  & 7.14  & 7.38 \\
        &         &         &         &               &               \\
Mean... & \nodata & \nodata & \nodata & 7.22$\pm$0.07 & 7.46$\pm$0.07 \\
        &         &         &         &               &               \\
Mg\,{\sc ii} $\lambda 4481.126$* &  8.864 & $+$0.749  &     &       &      \\
Mg\,{\sc ii} $\lambda 4481.150$* &  8.864 & $-$0.553  &     &       &      \\
Mg\,{\sc ii} $\lambda 4481.325$* &  8.864 & $+$0.594  &  51 & 5.77  & 5.88 \\
        &         &         &         &               &               \\
Al\,{\sc iii} $\lambda 4479.885$ & 20.781 & $+$0.900\tablenotemark{c}  &     &       &       \\
Al\,{\sc iii} $\lambda 4479.971$ & 20.781 & $+$1.020\tablenotemark{c}  &     &       &       \\
Al\,{\sc iii} $\lambda 4480.009$ & 20.781 & $-$0.530\tablenotemark{c}  & 20  &       & 5.11  \\
        &         &         &         &               &               \\
Al\,{\sc iii} $\lambda 4512.565$ & 17.808 & $+$0.410  & 21  &       & 5.22 \\
        &         &         &         &               &               \\
Al\,{\sc iii} $\lambda 4528.945$ & 17.818 & $-$0.290  &     &       &      \\
Al\,{\sc iii} $\lambda 4529.189$ & 17.818 & $+$0.660  & 33  &       & 5.17 \\
        &         &         &         &               &               \\
Al\,{\sc iii} $\lambda 5696.604$ & 15.642 & $+$0.230  & 21  &       & 4.96 \\
Al\,{\sc iii} $\lambda 5722.730$ & 15.642 & $-$0.070  & 11  &       & 4.95 \\
        &         &         &         &               &               \\
Mean... & \nodata & \nodata & \nodata &               & 5.08$\pm$0.12 \\
        &         &         &         &               &               \\
Si\,{\sc ii} $\lambda 5041.024$ & 10.066 & $+$0.029  &   8 &       & 6.11 \\
Si\,{\sc ii} $\lambda 5055.984$ & 10.074 & $+$0.523  &  15 &       & 5.90 \\
        &         &         &         &               &               \\
Mean... & \nodata & \nodata & \nodata &               & 6.01$\pm$0.15 \\
        &         &         &         &               &               \\
Si\,{\sc iii} $\lambda 4552.622$* & 19.016 & $+$0.292  &  70 & 6.11  & 5.93 \\
Si\,{\sc iii} $\lambda 4567.840$* & 19.016 & $+$0.068  &  46 & 6.03  & 5.85 \\
Si\,{\sc iii} $\lambda 4574.757$* & 19.016 & $-$0.409  &  20 & 6.01  & 5.84 \\
        &         &         &         &               &               \\
Si\,{\sc iii} $\lambda 4819.712$* & 25.982 & $+$0.937  &     &       &      \\
Si\,{\sc iii} $\lambda 4819.814$* & 25.982 & $-$0.354  &  12 & 5.75  & 5.99 \\
        &         &         &         &               &               \\
Si\,{\sc iii} $\lambda 4828.951$* & 25.987 & $+$0.937  &     &       &      \\
Si\,{\sc iii} $\lambda 4829.111$* & 25.980 & $-$0.354  &  16 & 5.90  & 6.15 \\
        &         &         &         &               &               \\
Si\,{\sc iii} $\lambda 5739.734$* & 19.722 & $-$0.096  &  23 & 6.16  & 5.95 \\
        &         &         &         &               &               \\
Mean... & \nodata & \nodata & \nodata & 5.99$\pm$0.15 & 5.95$\pm$0.11 \\
        &         &         &         &               &               \\
S\,{\sc ii} $\lambda 5432.797$ & 13.617 & $+$0.205  &  12 & 6.08  & 6.09 \\
        &         &         &         &               &               \\
S\,{\sc iii} $\lambda 4253.589$* & 18.244 & $+$0.107  &  30 & 6.30  & 5.71 \\
S\,{\sc iii} $\lambda 4284.979$* & 18.193 & $-$0.233  &  17 & 6.31  & 6.11 \\
S\,{\sc iii} $\lambda 4332.692$* & 18.188 & $-$0.564  &  $\leq$15 & $\leq$6.59  & $\leq$6.39 \\
        &         &         &         &               &               \\
Mean... & \nodata & \nodata & \nodata & 6.31$\pm$0.01 & 5.91$\pm$0.28 \\
        &         &         &         &               &               \\
Fe\,{\sc iii} $\lambda 4395.755$ &  8.256 & $-$2.595\tablenotemark{c}  &  $\leq$19  & $\leq$7.09  & $\leq$6.79 \\
Fe\,{\sc iii} $\lambda 4419.596$ &  8.241 & $-$2.218\tablenotemark{c}  &  $\leq$29  & $\leq$6.96  & $\leq$6.62 \\
Fe\,{\sc iii} $\lambda 4431.019$ &  8.248 & $-$2.572\tablenotemark{c}  &  $\leq$16  & $\leq$6.99  & $\leq$6.68 \\
Fe\,{\sc iii} $\lambda 5127.387$ &  8.659 & $-$2.218\tablenotemark{c}  &  $\leq$20  & $\leq$6.98  & $\leq$6.59 \\
Fe\,{\sc iii} $\lambda 5156.111$ &  8.641 & $-$2.018\tablenotemark{c}  &  $\leq$14  & $\leq$6.60  & $\leq$6.21 \\
        &         &         &         &               &               \\
Mean... & \nodata & \nodata & \nodata & $\leq$6.60$\pm$0.00 & $\leq$6.21$\pm$0.00 \\
\enddata
\tablenotetext{a}{($T_{\rm eff}$, $\log g$, $\xi$)=(22000, 3.45, 10.0)}
\tablenotetext{b}{($T_{\rm eff}$, $\log g$, $\xi$)=(21000, 3.35, 10.0)}
\tablenotetext{c}{Kurucz $gf$-value}
\tablenotetext{*}{The lines covered by the adopted model atom, including the
extended model atom with more levels, for the ions providing the ionization balance and some other ions}
\end{deluxetable}

\begin{deluxetable}{lccc}
\label{t:lines.uves}
\tablewidth{0pt}
\tablecolumns{4}
\tablecaption{Summary of V652\,Her's photospheric abundances}
\tablehead{
\colhead{Element} & \colhead{non-LTE} & \colhead{LTE} &
\colhead{Sun\tablenotemark{a}}}
\startdata
H   &  9.5 &  9.7 & 12.0 \\
He  & 11.5 & 11.5 & 10.9 \\
C   & 7.0 & 6.9 &  8.4 \\
N   & 8.7 & 8.9 &  7.8 \\
O   & 7.6 & 7.6 &  8.7 \\
Ne  & 8.1 & 8.6 &  7.9 \\
Mg  & 7.1 & 7.5 &  7.6 \\
Al  & \nodata & 6.4 &  6.5 \\
Si  & 7.4 & 7.4 &  7.5 \\
P  & \nodata & 5.5 &  5.4 \\
S  & 7.4 & 7.2 &  7.1 \\
Ar  & \nodata & 6.9 &  6.4 \\
Fe  & 7.1 & 7.3 &  7.5 \\
\enddata
\tablenotetext{a}{\citet{asplund09}}
\end{deluxetable}

\begin{deluxetable}{lccc}
\label{t:lines.uves}
\tablewidth{0pt}
\tablecolumns{4}
\tablecaption{Summary of HD\,144941's photospheric abundances}
\tablehead{
\colhead{Element} & \colhead{non-LTE} & \colhead{LTE} &
\colhead{Sun\tablenotemark{a}}}
\startdata
H   &  10.4 & 10.4 &  12.0 \\
He  &  11.5 & 11.5 &  10.9 \\
C   &  6.9  & 6.6  &   8.4 \\
N   &  6.4  & 6.7  &   7.8 \\
O   &  7.1  & 7.0  &   8.7 \\
Ne  &  7.2  & 7.5  &   7.9 \\
Mg  &  5.8  & 5.9  &   7.6 \\
Al  &  \nodata & 5.1 & 6.5 \\
Si  &  6.0  & 6.0  &   7.5 \\
S   &  6.2  & 6.0  &   7.1 \\
Fe  &  $\leq$6.6 & $\leq$6.2 & 7.5 \\
\enddata
\tablenotetext{a}{\citet{asplund09}}
\end{deluxetable}

\clearpage


\end{document}